\documentclass[pdflatex,sn-mathphys-num]{sn-jnl}%

\usepackage{geometry}
\usepackage{float}
\usepackage{graphicx}%
\usepackage{multirow}%
\usepackage{amsmath,amssymb,amsfonts}%
\usepackage{amsthm}%
\usepackage{mathrsfs}%
\usepackage[title]{appendix}%
\usepackage{adjustbox}
\usepackage{booktabs}
\usepackage{makecell}
\usepackage{xcolor}%
\usepackage{threeparttable}
\usepackage{textcomp}%
\usepackage{manyfoot}%
\usepackage{caption}
\usepackage{algorithm}%
\usepackage{algorithmicx}%
\usepackage{algpseudocode}
\usepackage{placeins}
\usepackage{listings}
\theoremstyle{thmstyleone}

\theoremstyle{thmstyletwo}%
\theoremstyle{thmstylethree}%
\usepackage{longtable}
\usepackage{booktabs}
\usepackage{array}
\newcolumntype{L}[1]{>{\raggedright\arraybackslash}p{#1}}
\usepackage{xr}
\begin{document}

\title[Uncertainty-aware emulation of maize traits]{Uncertainty-aware neural emulation reveals climate-resilient maize traits at scale}

\author[1]{\fnm{Mojdeh} \sur{Saadati}}\email{msaadati@iastate.edu}

\author[2]{\fnm{Juan} \sur{Panelo}}\email{jpanelo@ufl.edu}

\author[2]{\fnm{Gustavo} \sur{C. Visentini}}\email{visentinig@ufl.edu}
\author[3]{\fnm{Soumik} \sur{Sarkar}}\email{soumiks@iastate.edu}

\author*[2]{\fnm{Carlos} \sur{Messina}}\email{cmessina@ufl.edu}

\author*[3]{\fnm{Baskar} \sur{Ganapathysubramanian}}\email{baskarg@iastate.edu}

\affil[1]{\orgdiv{Department of Mathematics and Department of Computer Science}, \orgname{Iowa State University}, \orgaddress{\city{Ames}, \state{IA}, \country{USA}}}

\affil[2]{\orgdiv{Department of Horticultural Sciences}, \orgname{University of Florida}, \orgaddress{ \city{Gainesville}, \state{FL}, \country{USA}}}

\affil[3]{\orgdiv{Department of Mechanical Engineering, and Translational AI Center}, \orgname{Iowa State University}, \orgaddress{\city{Ames}, \state{IA}, \country{USA}}}

\abstract{
Climate change threatens maize yields worldwide, yet identifying which genetic traits sustain productivity under future conditions requires exploring an enormous space of genotype, soil, and climate combinations that process-based crop models are too costly to simulate at scale. Here we develop an integrated probabilistic emulation framework that transforms a mechanistic crop simulator (APSIM) into a fast, uncertainty-aware discovery engine. Built from two million simulations and coupled to a generative weather model that produces realistic climate sequences, the emulator reproduces thirteen maize growth outputs with high fidelity while returning calibrated confidence for every prediction, at a fraction of the original computational cost. Applying it across 100,000 trait configurations, six locations, and future climate projections extending to the year 2100, we identify 1,114 trait combinations that remain resilient under the 2100 future conditions and a stricter subset of 246 configurations that remain resilient across four future projection periods: 2040, 2060, 2080, and 2100. These resilient configurations occupy a distinct region of trait space, combining high radiation use efficiency with characteristic shifts in developmental and reproductive traits. Trait-importance analysis further identifies radiation use efficiency as the dominant driver of yield under future climate, while projected yields vary widely across locations and climate models, indicating strong spatial heterogeneity in the maize response to climate change.
}


\maketitle
\section{Introduction}\label{sec1}

Maize is the world's most-produced cereal, with annual production approaching 
1.2 billion tonnes and the fastest growth among the major grains, underpinning 
global food, feed, and bioenergy systems \cite{fao_food_outlook_2026_june}. Yet its 
productivity is increasingly threatened by climate change: under high-emission 
scenarios, mean end-of-century maize yields are projected to decline by roughly 
24\% as heat stress, drought, and shifting precipitation intensify across major 
growing regions \cite{jagermeyr2021climate}. Adapting to these pressures, 
whether through breeding, management, or both, requires understanding how maize 
responds across vast combinations of genetic traits, soil conditions, and 
future climate scenarios. Process-based crop models meet this need by 
mechanistically simulating crop development from genetic, soil, weather, and 
management inputs, allowing counterfactual exploration of conditions not 
yet observed.

Among available process-based crop models, APSIM 
stands out for its modular architecture, detailed representation of 
soil--crop--climate--management interactions, and extensive cross-environment 
validation \cite{mccown1996apsim, holzworth2014apsim, brown2018apsim}, making 
it a widely adopted tool for climate-risk analysis and farming-systems design 
and a compelling candidate for emulation: a machine-learning surrogate 
approximating APSIM outputs can retain much of its biophysical fidelity while 
dramatically reducing runtime \cite{johnston2023comparison}. However, despite 
their biophysical realism, these models remain computationally expensive: a 
single APSIM simulation requires approximately two minutes of compute time, 
making the millions of runs needed for large-scale trait discovery and climate 
exploration computationally prohibitive.
Machine-learning surrogates have emerged as a promising approach to alleviate 
this bottleneck by approximating the behavior of expensive mechanistic 
simulators at a fraction of the cost \cite{bocquet2023surrogate, 
pawar2022equation}. Neural emulators have demonstrated success across Earth 
system sciences, including climate dynamics, hydrology, and ocean 
modeling, enabling rapid approximation of numerical simulations while 
preserving key system behaviors \cite{gong2016multiobjective, 
zhang2018advances, jin2024gwsm4c}. In crop modeling, recent studies show that 
data-driven emulators can reproduce APSIM and DSSAT outputs across a range of 
applications, including chickpea yield prediction \cite{johnston2023comparison}, 
sugarcane calibration \cite{gunarathna2023emulator}, soil organic carbon 
estimation \cite{luo2019mapping}, continental-scale yield assessment 
\cite{vlachopoulos2026surrogate}, and deep learning hybrids for yield prediction 
\cite{shi2025deep}, with strong accuracy and substantial speedups 
\cite{ramankutty2013statistical, blanc2017statistical}.

Realizing this promise at the scale that agronomic discovery demands requires
more than a faster simulator.
Questions such as which trait combinations sustain yield under
end-of-century warming, how optimal nitrogen management shifts across
locations and emissions scenarios, or how genotype--environment
interactions evolve over decades each demand evaluating millions of
configurations, impractical with mechanistic simulation alone.
Yet existing crop emulators remain insufficient for discovery at this scale,
falling short in three respects.   \textit{First}, many rely on
simplified or temporally aggregated weather inputs and training data of
limited environmental diversity, so they neither capture the sequential
climate dynamics that drive phenological development and stress responses
nor generalize to novel conditions \cite{ramankutty2013statistical,
blanc2017statistical, johnston2023comparison}. \textit{Second}, most
emulators target yield as a single output instead of reproducing the full
multivariate behavior of the mechanistic model, limiting their utility for
mechanistic interpretation and multi-trait analysis
\cite{ramankutty2013statistical, blanc2017statistical, johnston2023comparison,
gunarathna2023emulator}. \textit{Third}, and most critically for
decision-making, most existing approaches produce deterministic predictions
without quantifying predictive uncertainty \cite{ramankutty2013statistical,
blanc2017statistical, johnston2023comparison, gunarathna2023emulator}. In
risk-sensitive applications such as climate adaptation planning and breeding
target identification, overconfident predictions can be more misleading than
no prediction at all, particularly when applied outside the training
distribution.

These three gaps define three requirements that a surrogate must meet, which
existing emulators satisfy only in part \cite{ramankutty2013statistical,
blanc2017statistical, johnston2023comparison, gunarathna2023emulator}: it must
(a) span the full space of genetic, soil, management, and climate variation,
including realistic sequential weather \cite{holzworth2014apsim, brown2018apsim};
(b) reproduce the mechanistic model's multivariate outputs, not yield alone; and
(c) attach calibrated uncertainty to every prediction, so that reliable
predictions can be separated from unreliable extrapolations.

Here we introduce an integrated probabilistic emulation framework for maize 
that addresses these challenges as a single system. Although its individual 
components draw on established techniques, their integration (large-scale 
mechanistic sampling, generative weather synthesis, multivariate neural 
emulation, and post-hoc uncertainty calibration) constitutes a pipeline that 
converts a slow mechanistic simulator into a calibrated, on-demand discovery 
engine. Specifically, we generate two million APSIM simulations using
low-discrepancy sampling of 21 independent variables and one derived variable,
for a total of 22 model inputs; couple these to convolutional weather representation models trained jointly on historical observations and downscaled future climate projections to generate synthetic weather sequences from the learned latent weather distribution; train a neural emulator that reproduces 13 crop-system outputs with 
high fidelity; and add a Stochastic Weight Averaging--Gaussian (SWAG) layer 
that provides calibrated predictive confidence without the cost of full 
Bayesian inference \cite{maddox2019simple}.

As a demonstration of what such a framework makes possible, we apply it to a combinatorially large space, spanning $100{,}000$ trait configurations, six locations, and future climate projections extending to the year 2100. We identify 1,114 maize trait combinations that consistently sustain high yield across the 2100 future conditions, and a stricter subset of 246 configurations that remains in the top-performing intersection across four future projection periods from 2040 to 2100, a scale of discovery impractical with the mechanistic model alone. We further show that these resilient configurations occupy a distinct region of trait space, characterized by high radiation use efficiency and shifts in developmental and reproductive traits, with trait-importance analysis identifying radiation use efficiency as the dominant driver of yield resilience under future climate. Projected yield responses also vary widely across locations and climate models, indicating strong spatial heterogeneity in climate response; climate change may thus reshape regional patterns of maize productivity.

Beyond crop science, this framework provides a general pathway for transforming 
computationally intensive mechanistic models into scalable, uncertainty-aware 
inference engines for Earth system and agricultural modeling.

\begin{figure}[!htbp]
    \makebox[\textwidth][c]{%
        \includegraphics[width=1.14\textwidth]{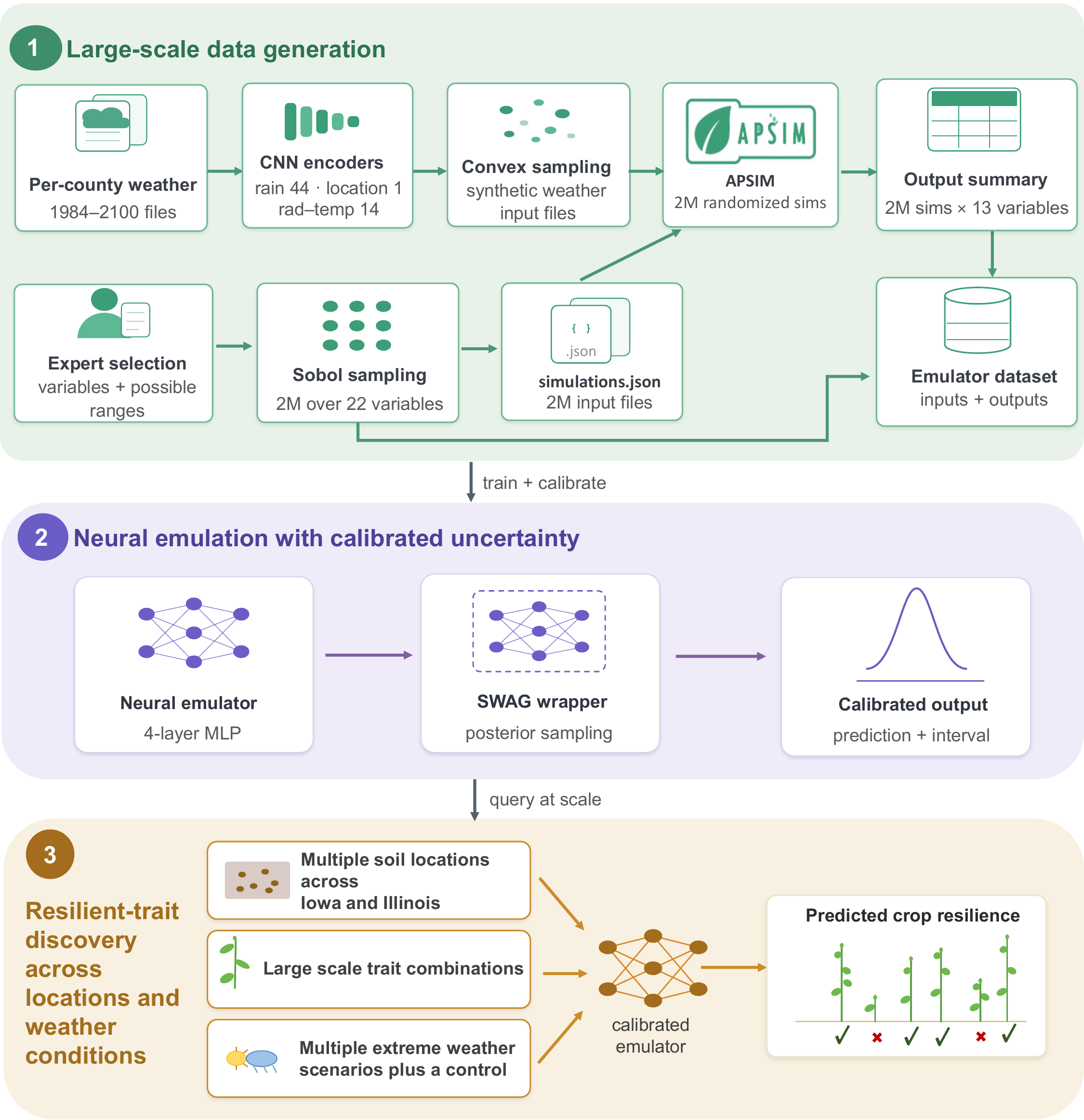}
    }
    \captionsetup{width=1.14\textwidth}
    \caption{Overview of the probabilistic neural emulation framework for APSIM 
    crop simulations. (1) Large-scale dataset generation via encoded weather 
    inputs and Sobol-sampled crop parameters. (2) Training a neural network 
    emulator with SWAG-based uncertainty estimation. (3) Example application: 
    identifying resilient crop traits under diverse conditions. The framework 
    can be queried for any large-scale genotype--environment--climate question 
    at negligible additional cost.}
    \label{fig:workflow}
\end{figure}

\section{Results}

Our framework transforms APSIM from a simulation-by-simulation tool into a
scalable, uncertainty-aware discovery engine organized into three stages
(Fig.~\ref{fig:workflow}). First, Sobol sampling is applied to 21 independently
varied genetic, environmental, and management variables, together with SWCON,
which is determined by the selected soil category, giving 22 total model
inputs. 
Historical observations and downscaled future climate projections are
encoded into a latent weather space, where new samples are drawn and decoded
to generate synthetic weather sequences. These sequences
are paired with the sampled inputs and evaluated with APSIM to produce two
million labeled simulations.

Second, a multivariate neural network is trained to emulate 13 APSIM outputs,
including phenological timing, integrated leaf area, biomass, grain yield and
size, grain-number response, above-ground and grain nitrogen accumulation,
canopy transpiration, and soil evaporation. A Stochastic Weight
Averaging--Gaussian (SWAG) layer provides calibrated predictive uncertainty.

Third, the trained emulator evaluates $100{,}000$ trait configurations across
48 future location--climate-model--scenario combinations (six locations $\times$
four climate models $\times$ two emissions scenarios), together with six separate
2020 baseline cases used for reference comparisons. The 1,114-configuration
resilience intersection is defined only across the 48 future cases. A stricter
temporal analysis subsequently evaluates the four future periods 2040, 2060,
2080, and 2100, producing 192 future location--climate-period cases in total.
We first validate the weather
generation model and APSIM emulator before applying the framework to
large-scale resilient-trait discovery.

\subsection*{Accurate reconstruction of weather dynamics via latent encoding}

The available historical and projected weather records provided 12,928 annual sequences, whereas the emulator analysis required a much larger set of weather realizations. Using only the downloaded records would provide relatively sparse coverage of plausible weather variability and increase dependence on particular weather realizations. We therefore generated additional synthetic weather data to expand the weather domain available for emulator training and improve the robustness of the downstream resilience analysis.

The weather generator uses a convolutional autoencoder to compress daily meteorological sequences into a compact latent space. The model was trained and evaluated using historical weather from 1984--2023 and downscaled future projections from 2025--2100 at five-year intervals. It reconstructs the four variables required by APSIM---solar radiation, maximum temperature, minimum temperature and rainfall---with high fidelity (correlation $>0.96$ and $R^2>0.92$ for all variables; Table~\ref{tab:ae_performance}). Correlations for radiation and temperature exceed 0.97, with negligible reconstruction bias. Rainfall amount, evaluated on wet days only, remains the most challenging variable (RMSE $=2.019$~mm~d$^{-1}$, $R^2=0.922$), reflecting its intermittent and highly variable behavior.

For rainfall occurrence, the model achieves an accuracy of 0.884 and an F1 score of 0.885. Its higher precision than recall (0.930 versus 0.844) indicates that predicted wet days are generally correct, although some observed rainfall events are missed. Together, these results show that the learned latent representation preserves the meteorological structure needed for synthetic-weather generation and downstream APSIM simulation.

Reconstruction accuracy alone, however, does not establish that the generated sequences cover the weather conditions represented in the archive. We therefore evaluated the 1.8 million synthetic weather sequences used for emulator training and compared their crop-relevant weather statistics with those of the archived weather data and the 48 future location--climate-model--scenario cases used for resilient-trait selection. As shown in Supplementary Fig.~\ref{fig:selected_weather_ecdf_coverage}, the synthetic distribution covers the archived weather domain, including the future cases. Although the future cases occupy a different part of the broader weather distribution, they remain within the overall encoder-training range for all four examined statistics.

\begin{table}[htbp]
\centering
\caption{Performance of the weather autoencoder in reconstructing key
meteorological variables on a held-out test set containing historical weather
and downscaled future climate projections. Correlation (Corr) denotes the
Pearson correlation coefficient between observed and reconstructed values.
Rainfall amount metrics are calculated on wet days only, while rainfall
occurrence is evaluated as a separate wet--dry classification task.}
\label{tab:ae_performance}
\small
\setlength{\tabcolsep}{8pt}
\renewcommand{\arraystretch}{1.2}
\begin{tabular}{lccccc}
\toprule
\multicolumn{6}{c}{\textbf{Continuous Reconstruction Metrics}} \\
\midrule
\textbf{Variable} & \textbf{$R^2$} & \textbf{Corr} & \textbf{Bias} & 
\textbf{RMSE} & \textbf{MAE} \\
\midrule
Radn (MJ m$^{-2}$ d$^{-1}$) & 0.9553 & 0.9774 &  0.005 & 1.620 & 1.200 \\
MaxT ($^\circ$C)             & 0.9787 & 0.9893 &  0.013 & 1.786 & 1.364 \\
MinT ($^\circ$C)             & 0.9725 & 0.9861 & -0.004 & 1.853 & 1.387 \\
Rain (mm d$^{-1}$)\footnotemark[1] & 0.9221 & 0.9610 & -0.259 & 2.019 & 1.098 \\
\midrule
\multicolumn{6}{c}{\textbf{Rain Occurrence (Wet/Dry Classification)}} \\
\midrule
\textbf{Variable} & \textbf{Accuracy} & \textbf{Precision} & \textbf{Recall} & 
\textbf{F1} & \\
\midrule
Rain (wet vs.\ dry) & 0.8841 & 0.9302 & 0.8437 & 0.8848 & \\
\bottomrule
\end{tabular}
\footnotetext[1]{Rain amount statistics are computed on wet days only 
(rain $> 0$ mm day$^{-1}$).}
\end{table}

\subsection*{Neural emulator achieves high predictive accuracy with calibrated uncertainty}

\begin{figure}[htbp]
\centering
\includegraphics[width=1.0\textwidth]{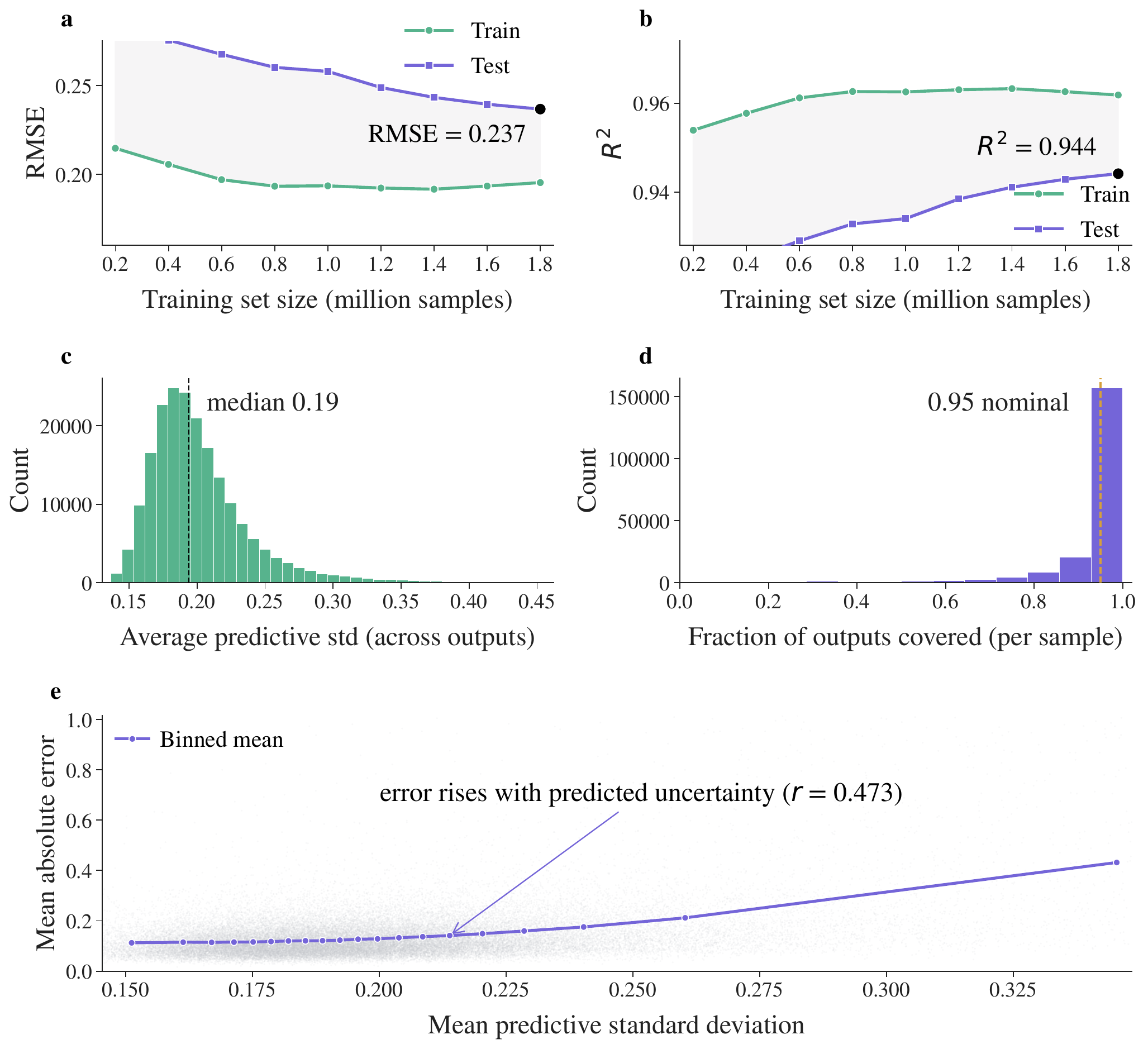}
\caption{Performance of the neural emulator and its SWAG uncertainty-quantification wrapper. 
(a,b) Learning curves for the deterministic neural emulator, showing training and test RMSE 
and $R^2$ as training-set size increases; improving test performance and a narrowing 
training--test gap indicate better generalization. (c--e) Uncertainty diagnostics for the 
SWAG wrapper applied to the trained emulator. (c) Distribution of average predictive standard 
deviation across outputs. (d) Per-sample coverage of the 95\% predictive intervals, with most 
samples near the nominal level. (e) Predictive uncertainty versus absolute prediction error, 
showing that higher uncertainty is associated with larger errors. Together, these results 
indicate well-calibrated uncertainty estimates that reflect emulator prediction difficulty.}
\label{fig:swag_performance_uncertainty}
\end{figure}

We next used the generated weather and APSIM outputs to train the neural emulator that supports the large-scale analysis. We first assessed its point-prediction performance using an independent held-out test set. Trained on 1.8 million samples, the emulator achieved a global normalized-space test $R^2=0.944$ and RMSE $=0.237$ across the 13 APSIM outputs. Test performance improved steadily as the training set increased: test $R^2$ rose from 0.918 with 0.2 million samples to 0.944 with 1.8 million samples, while normalized test RMSE declined from 0.288 to 0.237 (Fig.~\ref{fig:swag_performance_uncertainty}.a,b). In contrast, training $R^2$ remained between 0.954 and 0.963, and training RMSE remained between 0.192 and 0.215. The improvement in test performance as the dataset expanded, together with the relatively stable training performance, indicates improved generalization without evidence of substantial overfitting.

We then applied the SWAG uncertainty wrapper to the trained emulator. The resulting model achieved a global normalized-space $R^2=0.934$, RMSE $=0.257$, MAE $=0.148$, and MSE $=0.066$. The modest decrease in point-prediction accuracy relative to the deterministic model reflects the higher learning rate used during SWAG collection, which promotes exploration of a broader region of weight space and produces a more diverse predictive ensemble. This tradeoff is intentional: SWAG sacrifices a small amount of point-prediction accuracy to estimate sample-specific predictive uncertainty. Across outputs, MSE ranged from 0.030 for \textit{DAP\_to\_Maturity} to 0.106 for \textit{Maize.Grain.Size} (Table~\ref{tab:swag_regression_results}). The mean predictive standard deviation was 0.203, corresponding to an average 95\% predictive-interval width of 0.798 (Fig.~\ref{fig:swag_performance_uncertainty}.c,d).

The resulting intervals were well calibrated on the independent held-out test set, achieving 95.67\% empirical coverage compared with the nominal 95\% level. Per-output coverage ranged from 92.51\% to 98.07\%, indicating generally consistent calibration across APSIM outputs. Predictive uncertainty was also informative at the sample level: the mean predictive standard deviation correlated positively with both MAE ($r=0.473$) and MSE ($r=0.306$; Fig.~\ref{fig:swag_performance_uncertainty}.e). Thus, the SWAG wrapper provides accurate predictions together with uncertainty estimates that tend to increase for samples where the emulator makes larger errors.

\begin{table}[htbp]
\centering
\caption{Predictive accuracy and uncertainty-quantification performance of the SWAG wrapper applied to the trained neural emulator across 13 APSIM outputs. For each output, MSE and MAE are reported in normalized output space, with the corresponding physical-unit RMSE and MAE shown in parentheses. M.Var.\ denotes mean predictive variance, PI W.\ denotes mean 95\% predictive-interval width, and Cvrg denotes empirical coverage of the 95\% predictive interval. *Pearson Corr.\ denotes the sample-level Pearson correlation between the average predictive standard deviation and absolute prediction error (MAE).}
\label{tab:swag_performance}

\small
\renewcommand{\arraystretch}{1.12}


\setlength{\tabcolsep}{6.5pt}

\begin{tabular}{@{}lccccc@{}}
\toprule
\multicolumn{6}{c}{\textbf{Per-output SWAG performance}} \\
\midrule

\textbf{Output Variable} &
\textbf{MSE} &
\textbf{MAE} &
\textbf{M.Var.} &
\textbf{PI.W  } &
\textbf{Cvrg} \\

\midrule

DAP\_to\_Flowering
    & 0.074 (2.546)
    & 0.142 (1.325)
    & 5.364e-02
    & 0.867
    & 0.963 \\

  DAP\_to\_Maturity
    & 0.030 (2.934)
    & 0.097 (1.651)
    & 3.275e-02
    & 0.633
    & 0.977 \\

  DAP\_to\_Harvesting
    & 0.031 (3.043)
    & 0.103 (1.763)
    & 3.779e-02
    & 0.684
    & \textbf{0.981} \\

  Maize.LAI.Integral
    & 0.043 (16.083)
    & 0.126 (9.846)
    & 3.247e-02
    & 0.672
    & 0.959 \\

  Leaf.Transpiration.Cumulative
    & 0.056 (16.717)
    & 0.143 (10.103)
    & 3.987e-02
    & 0.760
    & 0.962 \\

  SoilWater.Es.Cumulative
    & 0.085 (11.158)
    & \textbf{0.191} (7.296)
    & \textbf{1.114e-01}
    & \textbf{1.279}
    & 0.979 \\

  Maize.AboveGround.N
    & 0.080 (1.283)
    & 0.164 (0.741)
    & 3.896e-02
    & 0.731
    & 0.925 \\

  Maize.AboveGround.Wt
    & 0.073 (102.896)
    & 0.152 (58.161)
    & 3.815e-02
    & 0.719
    & 0.946 \\

  Maize.Grain.N
    & 0.060 (0.895)
    & 0.152 (0.555)
    & 3.866e-02
    & 0.739
    & 0.937 \\

  Maize.Grain.NumberFunction
    & 0.073 (338.739)
    & 0.145 (181.196)
    & 4.452e-02
    & 0.783
    & 0.966 \\

  Maize.Grain.Size
    & \textbf{0.106} (0.020)
    & 0.182 (0.011)
    & 6.113e-02
    & 0.943
    & 0.957 \\

  Maize.Grain.Total.Wt
    & 0.071 (69.937)
    & 0.172 (44.901)
    & 4.259e-02
    & 0.772
    & 0.930 \\

  Maize.Total.Wt
    & 0.074 (109.388)
    & 0.154 (62.041)
    & 4.270e-02
    & 0.765
    & 0.955 \\

\bottomrule
\end{tabular}

\vspace{5pt}


\begin{tabular*}{\linewidth}
{@{\extracolsep{\fill}}cccccc@{}}

\toprule
\multicolumn{6}{c}{\textbf{Global SWAG wrapper metrics}} \\
\midrule

\textbf{MSE} &
\textbf{MAE} &
\textbf{RMSE} &
\textbf{$R^2$} &
\textbf{95\% Cvrg} &
\textbf{Pearson Corr.$^*$} \\

  0.066 &
  0.148 &
  0.257 &
  0.934 &
  0.957 &
  0.473\\

\bottomrule
\end{tabular*}
\label{tab:swag_regression_results}
\end{table}

\subsection*{Identifying resilient crop traits across climate scenarios and regions}

\subsubsection*{Yield responses exhibit strong spatial heterogeneity under climate change}

We simulated $100{,}000$ configurations of 12 genetic traits across six
  locations and eight climate projections derived from four global climate models
  under two emissions scenarios (SSP245 and SSP585) for the year 2100. The 2020 historical weather was used as the baseline comparator,
  providing a common reference for evaluating projected changes in climate and yield. An example of the contrast between the 2020 baseline and the
  2100 SSP585 projections is provided in Supplementary Figs.~\ref{fig:belleville_weather_comparison} and \ref{fig:belleville_rain_comparison}, which compare daily temperature and
  precipitation under SSP585 in 2100 with the 2020 baseline for Randolph County.
  The same genetic configurations were evaluated at every location and climate
  condition.
  Location-specific soil properties, including DUL, carbon, FInert, CN2Bare, and
  the derived SWCON profile, were retained for each site
  (Supplementary Table~\ref{tab:soil_location_properties}). Management practices,
  previous-crop residue, and initial soil water were held constant at a plant
  population of 8 plants m$^{-2}$, fertilizer application at sowing of
  190 kg ha$^{-1}$, a sowing date of 25 May, soybean residue (FOM = 0), and an
  initial soil-water value of 75\%. A 2020 baseline climate was additionally evaluated at each location as a reference for quantifying projected yield changes, resulting in 54 evaluated location--climate conditions in total.

Projected yield distributions under SSP585 show substantial variation across
locations and climate models (Fig.~\ref{fig:resilient_analysis}a). Median
yields under all future projections are lower than under the baseline climate,
although the magnitude of the reduction varies markedly among locations and
models. ACCESS-CM2 generally produces the strongest yield losses, while
IPSL-CM6A-LR produces particularly low yields at Osceola and Bremer; the
remaining models show more moderate but location-dependent responses. The
relative performance of locations is not preserved across climate models,
indicating that geographic advantage under one projected climate may not
translate to another. Variation in the width of the yield distributions further
shows that trait configurations differ in their sensitivity to future climate
conditions: some location--model combinations retain a broad range of
attainable yields, whereas others show narrower distributions concentrated at
lower productivity. This suggests that the potential for genotype-based
adaptation may itself be environment dependent, with greater opportunity for
selection in locations where genetic variation continues to produce a wider
yield response. Together, these results indicate that maize resilience cannot
be characterized by a single average climate effect; it emerges from the
combined influence of location, climate model, emissions pathway, and genotype.
Equivalent results for SSP245 and SSP585 across the four projection years are
provided in Supplementary Figs.~\ref{fig:violin_SSP245_4years} and
\ref{fig:violin_SSP585_4years}, respectively.

\subsubsection*{Future climate shifts yield toward radiation use efficiency}

\begin{figure}[!htbp]
\centering
\makebox[\linewidth][c]{%
\includegraphics[width=1.05\linewidth]{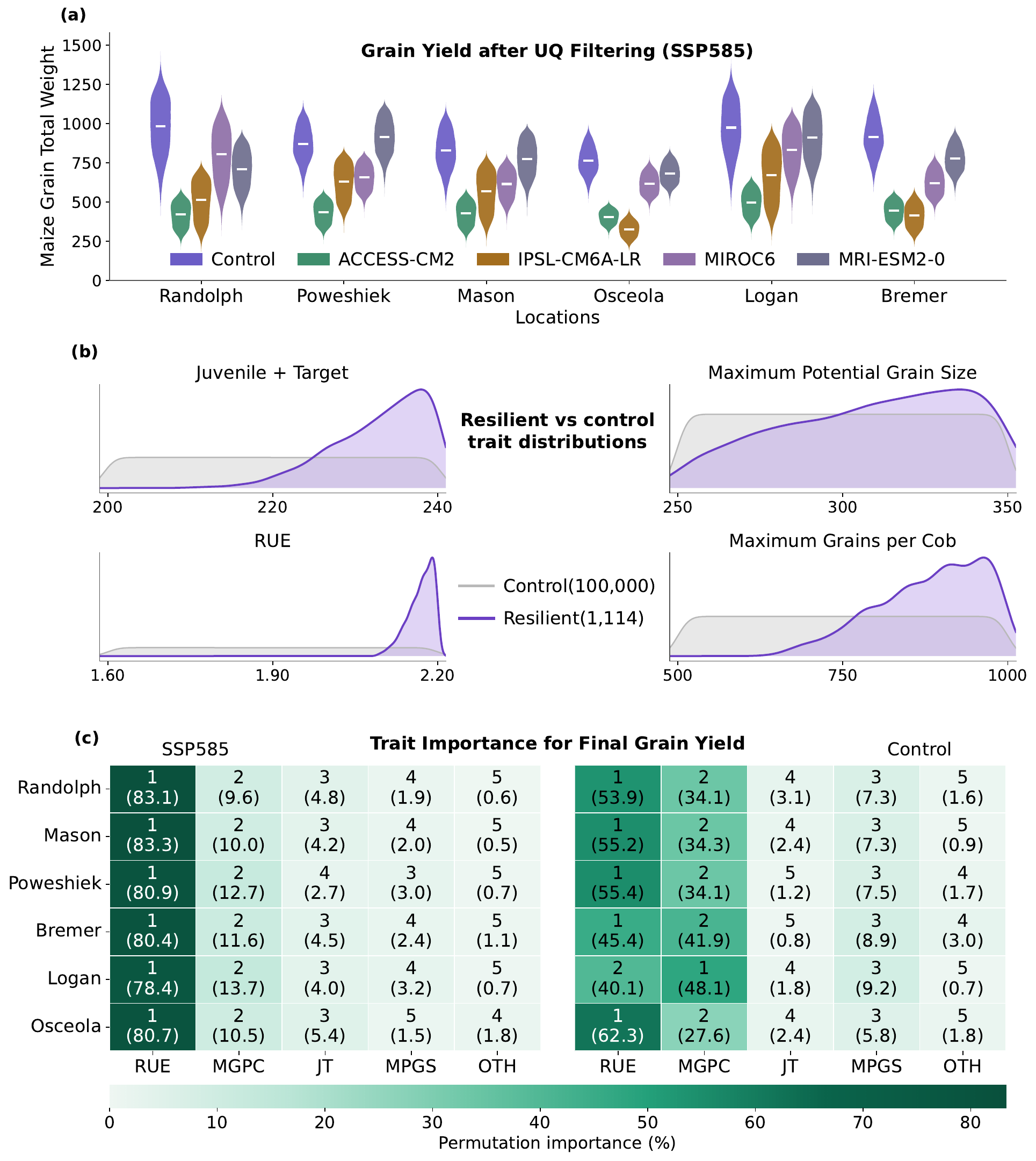}%
}
\caption{Future climate changes maize yield and the traits linked to yield stability.
\textbf{(a)} Yield responses under SSP585 differ across locations and climate models.
\textbf{(b)} High-yielding, stable configurations are associated with high RUE, later Juvenile+Target, larger potential grain size, and intermediate-to-high MaximumGrainsPerCob.
\textbf{(c)} Under future climate, RUE is the most important trait for yield, while baseline conditions show a more balanced contribution from RUE and reproductive traits. RUE, radiation use efficiency; MGPC, MaximumGrainsPerCob; JT, Juvenile+Target; MPGS, MaximumPotentialGrainSize; OTH, remaining traits. The corresponding multi-year SSP245 and SSP585 results are provided in Supplementary Figs.~\ref{fig:permutation_SSP245} and \ref{fig:permutation_SSP585}, respectively.}
\label{fig:resilient_analysis}
\end{figure}

Permutation importance analysis \cite{breiman2001random} using a random forest reveals a clear shift
in the traits governing final grain yield under future climate conditions
(Fig.~\ref{fig:resilient_analysis}c). Among the 12 evaluated traits, RUE, MaximumGrainsPerCob,
Juvenile+Target, and MaximumPotentialGrainSize consistently emerge among the highest-ranked contributors
across locations and climate conditions. Under SSP585, radiation use efficiency (RUE) becomes
consistently dominant at every location, accounting for approximately $78$--$83\%$ of total permutation
importance, while the other leading traits contribute only small fractions. Under the baseline climate,
importance is distributed more broadly across the top-ranked traits, particularly between RUE and
MaximumGrainsPerCob. RUE remains dominant at most locations, but its contribution decreases
substantially, while MaximumGrainsPerCob becomes much more influential and is the leading trait at
Logan. Juvenile+Target and MaximumPotentialGrainSize retain smaller but consistently high rankings
relative to the remaining eight traits. Permutation-importance results across SSP245 and SSP585 for the
four projection years (2040, 2060, 2080, and 2100) show persistent RUE dominance, with location- and
year-specific changes in the rankings of secondary traits; these results are provided in Supplementary
Figs.~\ref{fig:permutation_SSP245} and \ref{fig:permutation_SSP585}, respectively.

The results suggest that climate change narrows the range of physiological
  pathways associated with high yield. Under baseline conditions, yield variation
  is distributed across biomass accumulation, reproductive development, and
  phenology. Under SSP245 and SSP585, variation in yield is more strongly
  concentrated along the radiation-use-efficiency axis, while the contributions
  of reproductive and developmental traits become secondary. This pattern
  suggests that improving RUE may expand the capacity of maize to maintain
  productivity under future climates, with other traits providing complementary
  adaptation across locations (Supplementary Figs.~\ref{fig:permutation_SSP245}
  and \ref{fig:permutation_SSP585}).

\subsubsection*{Resilient configurations converge toward a distinct trait profile}

To identify trait combinations capable of maintaining high and reliable productivity across future environments, we applied yield and predictive-uncertainty criteria to the $100{,}000$ simulated genetic configurations. For each climate projection and location, predictive uncertainty was quantified using the coefficient of variation, $CV=\sigma/|\mu|$, where $\mu$ and $\sigma$ are the SWAG-predicted mean and standard deviation of grain yield. Configurations with $CV<0.4$ were ranked by predicted grain yield, and the top $5{,}000$ were retained. Intersecting the retained sets across all climate projections and six locations identified $1{,}114$ resilient configurations.

Compared with the full population, resilient configurations exhibit a distinct trait profile (Fig.~\ref{fig:resilient_analysis}b). RUE shows the strongest directional shift toward higher values, although its distribution declines before the upper bound of the expert-defined range. Juvenile+Target and MaximumPotentialGrainSize also shift toward larger values, while MaximumGrainsPerCob is concentrated in the upper portion of its range. The remaining traits retain broad distributions (Supplementary Fig.~\ref{fig:resilient_distribution_2100}). Similar trait-distribution patterns are observed in the stricter set of configurations that satisfy the resilience criteria across all four future projection years (Supplementary Fig.~\ref{fig:resilient_distribution_4y}). Together, these results associate resilience with high radiation-use efficiency and selective developmental and reproductive shifts, while allowing flexibility in the remaining traits.

Pairwise correlations within the 2100-resilient set are generally weak, with only a few moderate associations (Supplementary Fig.~\ref{fig:resilient_trait_correlation_2100}). PCA shows that these configurations occupy a structured region within the broader control trait space (Supplementary Fig.~\ref{fig:pca_resilient_control_comparison}).

\subsubsection*{Direct APSIM verification of resilient configurations}

Because the resilient configurations were identified with an emulator rather
than directly with APSIM, we verified the finalists against the mechanistic
simulator itself. Although the emulator reproduces APSIM behavior well overall,
approximation errors may remain for particular trait--environment combinations,
especially because the selected resilient traits and future weather conditions
occupy relatively sparse upper-tail regions of the broader distributions. These
errors could also affect the ordering of closely competing configurations, since
resilience was defined through repeated rank-based selection across environments.
Direct APSIM verification therefore tests both the numerical agreement of the
emulator and whether any environment-specific discrepancies propagate into the
final resilient set. We evaluated all 246 four-year resilient configurations
directly with APSIM across six locations, four climate models, two emissions
scenarios, and four future projection years, resulting in 47,232 matched
finalist--environment comparisons. APSIM and emulator predictions were compared
using absolute yield errors, normalized RMSE, and trait-level yield differences,
while ranking agreement was assessed using environment-specific absolute rank
distances. The overall normalized RMSE was 12.3\%, quantifying the magnitude of
the emulator's numerical approximation error across the matched cases. APSIM
predicted yields approximately 50~g~m$^{-2}$ higher than the
emulator on average, indicating a systematic offset rather than purely random
disagreement. This offset is modest relative to the range of yields represented
in the finalist--environment comparisons and is plausible given that the
verification focuses on high-performing trait combinations and challenging
future-weather conditions near the upper tails of the response distributions.
The mean and median absolute rank differences were 48 and 37 ranks,
respectively, among the 246 finalists, providing a direct assessment of whether
emulator errors altered the ordering of candidate configurations. Together,
these results support the overall robustness of the resilient-trait conclusions
while documenting the remaining emulator approximation error. The trait-level
yield differences, stratified prediction errors, and rank-disagreement
distributions are provided in
Supplementary Figs.~\ref{fig:apsim_emulator_trait_average_difference},
\ref{fig:apsim_emulator_normalized_rmse_strata}, and
\ref{fig:apsim_emulator_stratified_rank_difference}.

\section{Discussion}

This study suggests that future climate alters not only maize yield magnitude, but potentially also the physiological pathways through
  which genetic variation influences productivity. The strong dominance of radiation use efficiency (RUE) under SSP245 and SSP585
  indicates that future environments may increasingly constrain yield through biomass-conversion efficiency. Under baseline
  conditions, reproductive and developmental traits contribute more substantially, whereas under future climate their relative
  importance is reduced, although MaximumGrainsPerCob remains an important secondary contributor. This shift suggests that
  climate stress may narrow the number of physiological pathways capable of sustaining high yield, highlighting RUE as a dominant
  predictor of yield resilience within the evaluated trait space (Supplementary
  Figs.~\ref{fig:permutation_SSP245} and \ref{fig:permutation_SSP585}).

  Projected yield responses nevertheless remain strongly dependent on location and climate model. The relative performance of
  locations changes across projections, and the width of the yield distributions also differs among environments
  (Supplementary Figs.~\ref{fig:violin_SSP245_4years} and
  \ref{fig:violin_SSP585_4years}). This indicates
  that some future conditions preserve greater scope for genetic differentiation and selection, whereas others produce lower and
  more concentrated yield distributions. Adaptation therefore cannot be defined against a single expected climate trajectory;
  breeding strategies must account for both environmental heterogeneity and uncertainty in future climate projections.

  The resilient configurations provide a complementary interpretation. Their trait distributions do not support simultaneous
  maximization of all favorable traits. Instead, RUE converges strongly toward the upper end of its range, while
  MaximumGrainsPerCob, Juvenile+Target, and MaximumPotentialGrainSize also shift toward higher values but retain variation across
  their respective ranges (Supplementary Figs.~\ref{fig:resilient_distribution_2100} and
  \ref{fig:resilient_distribution_4y}). This pattern suggests a distinction between a core trait required for broad robustness and secondary
  traits that can vary within that high-RUE background. \textit{Resilience therefore corresponds to a constrained region of
  viable trait combinations, not a single optimal ideotype.}

This has direct implications for breeding. Breeding programs could prioritize traits associated with broad environmental
  robustness, particularly RUE, while adjusting reproductive and developmental traits to local conditions. The 1,114 resilient
  configurations therefore represent alternative pathways within a bounded trait space rather than fixed breeding prescriptions.
  Such a strategy may provide greater flexibility under heterogeneous future environments and reduce reliance on a single
  universally optimal genotype.

The framework developed here extends well beyond the specific question addressed in this study. By reducing simulation costs to near-negligible levels, the emulator makes an entire class of previously intractable questions directly accessible. For example, it enables systematic exploration of which trait combinations maximize water-use efficiency under drought, how optimal nitrogen management strategies vary across counties and emissions scenarios, which currently marginal soils may become productive under end-of-century warming, and how genotype-by-environment (G$\times$E) interactions evolve over time. Because each query costs a fraction of a mechanistic run, the emulator is well suited to serve as the biophysical engine within larger automated analysis systems. In recent work, it is embedded in an agentic decision-support framework that couples crop and economic models to answer natural-language policy questions about agricultural resilience \cite{waite2026ai}. The value of the emulator is greatest precisely in contexts where the scale of exploration exceeds what mechanistic simulation alone can support.

Several limitations bound these conclusions. The emulator inherits the structural assumptions of APSIM, so any systematic bias in the underlying process representations propagates into the trait rankings; a formal sensitivity analysis of how these assumptions influence identified resilience targets would strengthen confidence in them. The RUE shift and its permutation importance should therefore be interpreted as conclusions conditional on the expert-defined trait ranges and should not be extrapolated beyond that domain. Management was held constant to isolate genetic and climatic effects, leaving adaptive agronomic responses unmodeled, and resilience was defined by rank-based yield thresholds, so alternative definitions based on yield stability, economic return, or resource-use efficiency may surface complementary trait sets. Finally, the analysis is confined to Iowa and Illinois; although these capture the core of U.S. maize production and a representative range of soil conditions, generalization to other regions requires locally representative training data.

More broadly, this work illustrates that the boundary between mechanistic understanding and data-driven scalability is not fixed. Uncertainty-aware emulation provides a principled pathway across that boundary, enabling systematic exploration of high-dimensional design spaces while preserving interpretability and reliability. This approach is not limited to crop modeling; similar trade-offs between model fidelity and computational tractability constrain large-scale analysis across hydrology, climate dynamics, and Earth system science. It also changes how a mechanistic model is used: APSIM becomes a standing query engine that can be interrogated at scale as new questions arise, new climate projections become available, or new breeding targets are proposed. Just as genomic databases transformed which genetic questions could be asked, scalable emulation expands the space of solvable crop-modeling problems. As climate pressures intensify, such frameworks may become standard tools for advancing scientific understanding and accelerating the translation of simulation insight into actionable breeding and management strategies. Beyond the 1114 resilient configurations identified here, the reusable framework is this work's central contribution.

\section{Materials and Methods}

\subsection{Study Design and Overview}
The central goal of this study was to develop a surrogate modeling framework that makes 
large-scale scientific exploration with mechanistic crop models computationally feasible. 
By approximating APSIM at orders-of-magnitude lower computational cost, the framework 
enables hundreds of thousands of trait configurations to be evaluated across multiple 
locations, climate models, emissions scenarios, and environmental conditions within a 
practical amount of time. To construct the emulator, we generated a large-scale dataset of 
APSIM simulations spanning diverse genetic, soil, management, and weather conditions and 
trained a neural network to approximate the mechanistic model. Weather inputs were first 
encoded into compact latent representations to balance their dimensionality against the 
other model inputs, and a post-hoc uncertainty quantification layer was added to distinguish 
reliable predictions from uncertain extrapolations, enabling robust downstream analysis.

We applied this framework to the scientific question of how projected climate change may 
reshape maize yield distributions across diverse genetic, soil, and management conditions. 
Climate projections were obtained from the NASA NEX-GDDP-CMIP6 dataset \cite{thrasher2022nasa}, comprising four
global climate models (IPSL-CM6A-LR, ACCESS-CM2, MIROC6, and MRI-ESM2-0) under two
emissions scenarios (SSP245 and SSP585) for the year 2100 \cite{riahi2017shared}. Iowa and Illinois were selected 
as the study domain for their importance in U.S. maize production and their diversity of 
soil environments. Evaluating this question directly with the mechanistic model would require 
approximately $5.4 \times 10^{6}$ APSIM runs ($10^{5}$ trait configurations $\times$ 
[4 climate models $\times$ 2 emissions scenarios $+$ 1 baseline] $\times$ 6 locations); 
at approximately two minutes of compute per simulation, this amounts to about 
180{,}000 CPU-hours, making such an analysis prohibitive for routine exploration.

Using the emulator, we evaluated the full set of 100{,}000 trait configurations across all
54 conditions: 48 future location--climate-model--scenario cases and six 2020 baseline
cases. The 2020 baseline cases were used for comparison only and were not included in
the resilience intersection. A trait configuration was defined as resilient if
its predicted yield ranked within the top 5{,}000 configurations for every 
future location--climate-model--emissions-scenario combination tested, subject to a minimum
prediction confidence threshold. Prediction confidence was assessed using the coefficient
of variation (CV) of the SWAG ensemble outputs: for each location--climate
combination, configurations were retained only if CV $< 0.4$, excluding
predictions with high uncertainty relative to expected yield. The 2100 analysis therefore
used 48 future cases, whereas the stricter four-period analysis used 192 future cases
spanning 2040, 2060, 2080, and 2100.
Any averages reported for these conditions were used only for descriptive summaries and
did not determine membership in the resilient intersection.
\subsection{Dataset Creation for APSIM Emulator}

\noindent\textbf{Emulator input and output variables.}
Domain experts identified 21 independently sampled genetic, environmental,
and management variables, together with SWCON, which was determined by the
selected soil-hydraulic category, giving 22 total emulator inputs. The 13
output variables capture key aspects of crop development and resource use,
including phenological timing, integrated canopy leaf area, biomass, grain
yield and size, grain-number response, above-ground and grain nitrogen
accumulation, canopy transpiration, and soil evaporation (Fig.~\ref{fig:workflow}a). Soil hydraulic
properties were represented by four categories---sandy, loam,
silt loam, and clay loam---to limit combinatorial
complexity while retaining variation in soil-water behavior. The complete
input definitions and ranges are provided in Supplementary
Table~\ref{tab:variables_defs_ranges}, and the selected emulator outputs in
Supplementary Table~\ref{tab:apsim_outputs}.

Domain experts also specified the permissible range of each variable a priori;
all subsequent Sobol sampling, emulator evaluation, resilient-set selection,
and permutation-importance analyses were therefore conditional on this
scientifically defined trait domain.

\textbf{Sampling and simulation generation.}
We generated two million samples using a Sobol low-discrepancy sequence, which covers
high-dimensional parameter spaces more uniformly than random sampling at equivalent sample 
sizes \cite{sobol1967distribution}. For each sample, a corresponding APSIM configuration file was generated by modifying 
a base maize template (Fig.~\ref{fig:workflow}a).

\textbf{Large-scale simulation execution.}
Simulations were executed on the Frontera supercomputer at the Texas Advanced Computing 
Center, partitioned into 200 batches of 10{,}000 runs each, with 36 CPU cores per node. 
Each simulation produced daily time-series outputs, which were aggregated into the summary 
variables used for emulator training.

\subsection{Weather Representation and Synthetic Generation}
Historical daily weather for 1984--2023 was obtained for all counties in Iowa and 
Illinois, and future projections for the same region were drawn from the NASA 
NEX-GDDP-CMIP6 dataset (four global climate models---ACCESS-CM2, IPSL-CM6A-LR, 
MIROC6, and MRI-ESM2-0---under two emissions scenarios, SSP245 and SSP585, at 
five-year intervals from 2025 to 2100) \cite{thrasher2022nasa}. Each annual sequence comprised 365 daily 
values of solar radiation, maximum temperature, minimum temperature, and rainfall. 
For comparison with future projections, the year 2020 was selected as the representative historical
baseline because its climatology most closely matched the median over 2020--2025 (Fig.~\ref{fig:workflow}a).
This baseline was used only as a reference condition and was not included in either the 2100 or four-period
resilience intersection. Future
records were sampled to match the historical train/test counts without overlap, 
yielding a balanced historical--future dataset that captures both observed variability 
and changes expected under future emissions.

Because crop responses depend on the temporal structure and joint behavior of weather 
variables, we trained convolutional encoder--decoder models on this combined corpus, 
using separate models for radiation--temperature and rainfall to accommodate their 
distinct statistics. The radiation--temperature model jointly encoded daily solar 
radiation and maximum and minimum temperature into a 14-dimensional latent vector, 
while a separate probabilistic rainfall model used a 44-dimensional vector; appending 
latitude gave a 59-dimensional weather representation for sampling. Synthetic sequences 
were generated as convex combinations of nearby latent encodings: latent variables were 
standardized and latitude upweighted to favor geographically consistent neighbors, then, 
for each sample, several vectors from a local $k$-nearest-neighbor neighborhood were 
mixed with random convex weights and decoded into a complete 365-day meteorological 
sequence.

\subsection{Weather Model Architecture and Training}

Weather was represented using two separate convolutional encoder--decoder architectures 
designed to capture temporal dependencies in daily meteorological sequences 
(Fig.~\ref{fig:workflow}a). The first model jointly encoded solar radiation, maximum temperature, and minimum temperature  as a three-channel daily sequence, allowing the network to learn both 
within-variable temporal structure and dependencies among temperature and radiation. 
These three variables were standardized using training-set statistics before being 
passed to the encoder--decoder. A second, independently trained architecture was used 
specifically for rainfall, which was retained in its original physical units because of 
its highly zero-inflated and skewed distribution.

The rainfall model employed separate components for precipitation occurrence and 
intensity. Rainfall occurrence was represented through a wet-day probability, whereas 
rainfall amount on wet days was modeled in log space using a Gaussian likelihood with 
predicted mean and variance. Extreme precipitation events were additionally upweighted 
during training to improve representation of the upper tail of the rainfall distribution. 
The two models used distinct latent representations and optimization settings, and their 
latent features were subsequently combined for use by the downstream emulator. Detailed 
architectural configurations and training hyperparameters are provided in Supplementary
Table~\ref{tab:S_hyperparams}.
\subsection{Neural Emulator Training}

To approximate APSIM outputs, we trained a fully connected neural network on the simulated
dataset (Fig.~\ref{fig:workflow}b). Inputs and outputs were standardized using training-set statistics. The network
comprised four fully connected hidden layers (256--256--256--128 units) with SiLU
activations, followed by a linear output layer. The dataset was
partitioned into 1.8 million training samples and 0.2 million held-out test samples prior
to standardization, and performance was evaluated on the held-out set using standard
regression metrics. To determine the appropriate dataset size, we conducted a scaling
analysis of performance as a function of training-set size, identifying a point of
diminishing returns that balanced predictive accuracy against computational cost. Missing values in \verb|DAP_to_Flowering|, \verb|DAP_to_Maturity|, and
\verb|DAP_to_Harvesting| were excluded from the loss and metric calculations using
binary masks. Training procedures and hyperparameter configurations are provided in
Supplementary Table~\ref{tab:S_hyperparams}.

The held-out test dataset was constructed as a synthetic evaluation set rather
than as a collection of directly observed location--weather cases. Weather data
were drawn from multiple counties across Iowa and Illinois and from both
historical and future periods, including different climate scenarios. These
diverse weather sources were used to construct synthetic weather inputs through
convex sampling in the learned weather representation, thereby broadening the
emulator's training and evaluation domain. Soil and management variables were
generated independently through Sobol sampling over their prescribed parameter
ranges. Consequently, each test sample represents a synthetic combination
designed to support a robust emulator; it is not directly associated with a
particular county, observed weather year, or uniquely defined emissions-scenario
trajectory. The location and climate labels therefore do not define independent
real-world strata, and the sampled soil and management variables are not
physically tied to the corresponding weather labels. For this reason,
location-, year-, and scenario-specific SWAG coverage would not provide a
scientifically interpretable assessment of spatial generalization or uncertainty
propagation. We consequently report global empirical coverage on the independent
held-out test set as the primary calibration measure and do not treat arbitrary
stratification of this synthetic dataset as additional evidence of model
calibration.

\subsection{Uncertainty Quantification}

To quantify predictive uncertainty, we adopted a post-hoc approach based on Stochastic 
Weight Averaging--Gaussian (SWAG) (Fig.~\ref{fig:workflow}b), which approximates a posterior 
distribution over model parameters from the trajectory of stochastic gradient descent during 
fine-tuning. Following initial deterministic training, the pretrained emulator was fine-tuned 
with SGD for 40 epochs. SWAG statistics were accumulated throughout the fine-tuning phase, 
with one model snapshot collected after each full epoch, yielding 40 parameter snapshots 
used to estimate the mean and low-rank covariance structure of the approximate posterior. 
Thirty weight samples were subsequently drawn from the resulting approximate posterior to 
construct a predictive ensemble, from which predictive means and uncertainties were estimated 
and used for uncertainty-based filtering. Full implementation details are provided in
Supplementary Table~\ref{tab:S_hyperparams}.
\subsection*{Funding}
This work was supported by the AI Institute for Resilient Agriculture (USDA-NIFA \#2021-67021-35329), COALESCE: COntext Aware LEarning for Sustainable CybEr-Agricultural Systems (NSF CPS Frontier \#1954556).

\subsection*{Author Contributions} 

B.G., C.M., and S.S. conceived and designed the project. 
M.S. developed the overall framework, generated the large-scale simulation dataset, and implemented the neural emulator and weather encoder--decoder models. 
M.S. performed the computational experiments, uncertainty quantification, and primary data analysis. 
M.S. and B.G. jointly designed the experimental setup for the final analysis and interpreted the results. 
C.M., J.P., and G.V. contributed to the selection of emulator variables and provided domain expertise. 
M.S. wrote the initial draft of the manuscript. 
All authors contributed to editing, reviewing, and approving the final version of the manuscript.

\subsection*{Conflicts of Interest}
The authors declare that there are no competing interests
associated with this work.

\bibliography{sn-bibliography}

\end{document}


\section*{Supplementary Materials}

\section{APSIM Template Configuration and Emulator Variable Selection}

\subsection{Simulation Template Modifications}

Before generating the simulation ensemble, several fixed settings in the
original APSIM template were modified to better represent the intended maize
production system and improve simulation robustness.

\medskip
\noindent\textbf{Maize cultivar.}
The default cultivar was changed from \texttt{P1197} to
\texttt{Pioneer\_3153} so that the genetic parameters could be
assigned directly without being overwritten by cultivar-default values.
This also provided a consistent baseline for evaluating variation in
phenology, biomass accumulation, and yield.

\medskip
\noindent\textbf{Sowing soil-water requirement.}
The minimum extractable soil water required for sowing was reduced from 50 to
30. This modification increased the likelihood that planting would occur
within the prescribed sowing window across the broad range of soil and weather
conditions represented in the simulation ensemble, while still requiring
sufficient soil water for crop establishment.

\medskip
\noindent\textbf{Sowing rainfall requirement.}
The minimum rainfall required for sowing was reduced from 25 to 10~mm, and the
rainfall accumulation period was shortened from 7 to 5~days. These changes
prevented excessively restrictive rainfall criteria from delaying or
preventing planting under otherwise suitable conditions.

\medskip
\noindent\textbf{Initial soil nitrate.}
The default uniform NO$_3$ profile of 1.0~ppm in all seven soil
layers was replaced with a depth-dependent profile of
10, 7, 4, 3, 2, 1, and 0.5~ppm. This declining
profile provided greater mineral-nitrogen availability near the soil surface,
where early root activity and residue decomposition are concentrated, while
maintaining lower concentrations in deeper layers.

\medskip
\noindent\textbf{Initial soil ammonium.}
The default uniform NH$_4$ profile of 0.1 in all seven soil layers was replaced
with values of 0.5, 0.4, 0.3, 0.2, 0.15, 0.1, and 0.05. This adjustment
provided a small but depth-dependent ammonium pool and reduced the risk of
mineral-nitrogen depletion during pre-sowing organic-matter decomposition.

\medskip
\noindent\textbf{Surface-residue C:N ratio.}
The initial surface-residue C:N ratio was reduced from 100 to 70, while the
initial residue mass was retained at 500~kg\,ha$^{-1}$. This change represented
a less nitrogen-poor residue pool and reduced excessive nitrogen
immobilization during residue decomposition, thereby improving simulation
stability before fertilizer application.

\medskip
After establishing this common simulation template, the input variables
were varied across the simulation ensemble and the corresponding APSIM
responses were extracted for emulator development.
\subsection{Input and Output Variable Selection}
The input and output variables were selected in consultation with domain
experts. For the input variables, experts also defined the feasible ranges
from which values were sampled and assigned in each simulation. Parameters
that were neither sampled nor explicitly modified in the simulation template
were retained at their original values.
Table~\ref{tab:variables_defs_ranges} summarizes the sampled input variables
and their feasible ranges, and Table~\ref{tab:apsim_outputs} lists the APSIM
response variables used as emulator outputs.

\setlength{\LTleft}{0pt}
\setlength{\LTright}{0pt}
\setlength{\tabcolsep}{5pt}
\renewcommand{\arraystretch}{1.18}

\begin{center}
\begin{minipage}{\textwidth}

\small
\setlength{\tabcolsep}{5pt}
\renewcommand{\arraystretch}{1.2}

\captionof{table}{
Emulator output variables extracted from APSIM maize simulations.
Phenological variables are expressed as days after planting (DAP).
Biomass and nitrogen variables are recorded at the EndGrainFill stage,
while water-use variables are accumulated from sowing through EndGrainFill.
}
\label{tab:apsim_outputs}

\centering
\begin{tabular}{p{4.2cm} p{8.0cm}}
\toprule
\textbf{Variable} & \textbf{Description} \\
\midrule

DAP\_to\_Flowering $(d)$ &
Number of days after planting from sowing to the first occurrence of
flowering. \\

DAP\_to\_Maturity $(d)$ &
Number of days after planting from sowing to the first occurrence of
maturity; EndGrainFill is used when maturity is absent. \\

DAP\_to\_Harvesting $(d)$ &
Number of days after planting from sowing to the first occurrence of
the harvest-ripe stage; EndGrainFill is used when harvest-ripe is absent. \\

LAI.Integral $(\mathrm{LAI}\,d)$ &
Time-integrated leaf area index, calculated by trapezoidal integration
of daily LAI values from sowing through EndGrainFill. \\

AboveGround.Wt $(kg\,ha^{-1})$ &
Above-ground dry biomass at EndGrainFill. \\

Grain.Size $(g\,grain^{-1})$ &
Average dry mass of an individual grain at EndGrainFill. \\

Grain.Total.Wt $(g\,m^{-2})$ &
Total grain dry weight, representing grain yield, at EndGrainFill. \\

Total.Wt $(g\,m^{-2})$ &
Total plant dry biomass at EndGrainFill. \\

Grain.N $(g\,m^{-2})$ &
Nitrogen accumulated in grain at EndGrainFill. \\

AboveGround.N $(kg\,ha^{-1})$ &
Nitrogen accumulated in above-ground biomass at EndGrainFill. \\

Leaf.Transpiration $(mm)$ &
Cumulative maize canopy transpiration from sowing through EndGrainFill,
representing total plant water use. \\

SoilWater.Es $(mm)$ &
Cumulative soil evaporation from sowing through EndGrainFill,
representing direct water loss from the soil surface. \\

Grain.NumberFunction $(-)$ &
Dimensionless APSIM scaling factor controlling potential grain number
during grain set, recorded at EndGrainFill. \\

\bottomrule
\end{tabular}

\end{minipage}
\end{center}

\begin{center}
\begin{minipage}{\textwidth}

\small
\setlength{\tabcolsep}{5pt}
\renewcommand{\arraystretch}{1.18}

\captionof{table}{
Sampled and derived input variables, definitions, and ranges used in the
APSIM simulation design.
}
\label{tab:variables_defs_ranges}

\centering
\begin{tabular}{@{}
    p{0.050\textwidth}
    p{0.28\textwidth}
    p{0.45\textwidth}
    p{0.15\textwidth}
@{}}

\toprule
\textbf{\#}
& \textbf{Variable}
& \textbf{Definition}
& \hspace{-0.20in}\textbf{Range or values} \\
\midrule

1-G
& ShootLag ($^\circ$C\,day)
& Thermal-time delay associated with shoot emergence.
& 45--65\textsuperscript{\ensuremath{\dagger}} \\

2-G
& ShootRate ($-$)
& Rate parameter controlling shoot emergence.
& 0.4--0.8 \\

3-G
& JuvenileTarget ($^\circ$C\,day)
& Thermal-time target for the juvenile developmental phase (denoted Juvenile+Target in the main text).
& 200--240\textsuperscript{\ensuremath{\dagger}} \\

4-G
& FloweringToGrain \newline FillingTarget ($^\circ$C\,day)
& Thermal-time target from flowering to the beginning of grain filling.
& 100--200\textsuperscript{\ensuremath{\dagger}} \\

5-G
& GrainFillingTarget ($^\circ$C\,day)
& Thermal-time target for the grain-filling phase.
& 600--800\textsuperscript{\ensuremath{\dagger}} \\

6-G
& MaximumPotential\newline GrainSize
  (mg\,grain$^{-1}$)
& Maximum potential mass of an individual grain. The sampled value is
divided by 1000 before being supplied to APSIM.
& 250--350\textsuperscript{\ensuremath{\dagger}} \\

7-G
& MaximumGrainsPerCob
  (grains\,cob$^{-1}$)
& Maximum potential number of grains per cob.
& 500--1000\textsuperscript{\ensuremath{\dagger}} \\

8-G
& FinalNconc (kg\,kg$^{-1}$)
& Final nitrogen concentration of grain.
& 0.0067--0.016 \\

9-G
& TemperatureFactor1 ($^\circ$C)
& Lower intermediate temperature point in the temperature-response curve.
& 5--12\textsuperscript{\ensuremath{\dagger}} \\

10-G
& TemperatureFactor2 ($^\circ$C)
& Upper intermediate temperature point in the temperature-response curve.
& 20--27\textsuperscript{\ensuremath{\dagger}} \\

11-G
& PotentialExtinctionCoeff ($-$)
& Intermediate canopy light-extinction coefficient used in the
extinction-coefficient response curve.
& 0.3--0.5 \\

12-G
& RUE (g\,MJ$^{-1}$)
& Radiation-use efficiency.
& 1.6--2.2 \\

13-E
& DUL category ($-$)
& Soil hydraulic category determining the DUL, LL15, SAT, and SWCON
profiles.
& 0, 1, 2, 3%
\textsuperscript{\ensuremath{\ddagger}} \\

14-E
& Carbon ($-$)
& Scaling parameter used to construct the soil organic-carbon profile.
The resulting top-layer carbon concentration ranges from approximately
1\% to 5\%.
& 0.01--0.05 \\

15-E
& InitialValues (\%)
& Initial soil-water position between LL15 and DUL, expressed as a
percentage of the interval from LL15 to DUL.
& 50--100%
\textsuperscript{\ensuremath{\dagger}} \\

16-E
& FInert category ($-$)
& Category determining the inert fraction of soil organic matter in all
soil layers.
& 0, 1, 2 \mbox{for} 0.25, 0.50, 0.75 \\

17-E
& CN2Bare ($-$)
& USDA runoff curve number for bare soil.
& 60--100%
\textsuperscript{\ensuremath{\dagger}} \\

18-E
& SWCON (day$^{-1}$)
& Soil-water conductivity factor derived from the selected DUL category.
& 0.55, 0.40, 0.25, or 0.12%
\textsuperscript{*} \\

19-E
& FOM category ($-$)
& Previous-crop category determining the fresh organic-matter profile.
& 0, 1 for \newline soybean,maize%
\textsuperscript{\ensuremath{\ddagger}} \\

20-M
& Population (plants\,m$^{-2}$)
& Maize plant population density.
& 5--12%
\textsuperscript{\ensuremath{\dagger}} \\

21-M
& StartDate (\textit{date})
& Beginning of the sowing window, represented as an integer day offset
from April 15. The end of the sowing window is fixed at June 20.
& April 15 - May 25%
\textsuperscript{\ensuremath{\dagger}} \\

22-M
& FertilizeAtSowing
  (kg\,ha$^{-1}$)
& Fertilizer nitrogen applied at sowing.
& 30--300%
\textsuperscript{\ensuremath{\dagger}} \\

\midrule

\multicolumn{4}{@{}p{0.96\textwidth}@{}}{%
\footnotesize
\textit{Notes.}
G, E, and M denote genetic, environmental, and management variables,
respectively. The 21 variables other than SWCON were independently
sampled, whereas SWCON was derived from the DUL category.

\textsuperscript{\ensuremath{\dagger}}
Integer-valued variable produced by applying the floor operation after
Sobol sampling.

\textsuperscript{\ensuremath{\ddagger}}
Categorical variable.

\textsuperscript{*}
Derived variable.

DUL categories correspond to:
0: sandy or sandy loam,
1: loam,
2: silt loam, and
3: silty clay loam or clay loam.

The corresponding
$(\mathrm{DUL}, \mathrm{LL15}, \mathrm{SAT}, \mathrm{SWCON})$
values are
$(0.22, 0.09, 0.40, 0.55)$,
$(0.29, 0.13, 0.44, 0.40)$,
$(0.35, 0.17, 0.48, 0.25)$, and
$(0.40, 0.20, 0.52, 0.12)$, respectively.

For the FOM category, the first two soil-layer values are
1000 kg\,ha$^{-1}$ for soybean residue and
1500 kg\,ha$^{-1}$ for maize residue.
} \\

\bottomrule

\end{tabular}

\end{minipage}
\end{center}
\section{Hyperparameter tuning}
This section reports the complete hyperparameter configurations for all three model components developed in this study: the weather autoencoder (separate architectures for temperature--radiation and rainfall), the fully connected neural network emulator, and the SWAG uncertainty quantification wrapper. Parameters governing optimizer settings, regularization strength, learning rate scheduling, and architectural dimensions are consolidated here to support reproducibility.
\begin{table}[h]
\centering
\caption{Hyperparameter settings for all model components.}
\label{tab:S_hyperparams}

\small
\setlength{\tabcolsep}{8pt}
\renewcommand{\arraystretch}{1.2}

\begin{tabular}{llr}
\toprule
\textbf{Component} & \textbf{Parameter} & \textbf{Value} \\
\midrule

\multirow{6}{*}{\parbox{4.1cm}{Temperature--Radiation Autoencoder}}
  & Learning rate        & $1 \times 10^{-3}$ \\
  & Weight decay         & $3 \times 10^{-4}$ \\
  & Latent dimension     & 14 \\
  & Batch size           & 96 \\
  & Max epochs           & 500 \\
  & Reconstruction loss  & Smooth L1 ($\beta = 0.5$) \\

\midrule

\multirow{6}{*}{Rainfall Autoencoder}
  & Learning rate        & $5 \times 10^{-4}$ \\
  & Weight decay         & $1 \times 10^{-4}$ \\
  & Latent dimension     & 44 \\
  & Batch size           & 96 \\
  & Max epochs           & 500 \\
  & Training loss        & BCE + wet-day Gaussian NLL \\

\midrule

\multirow{7}{*}{Neural Emulator}
  & Optimizer            & AdamW \\
  & Learning rate        & $3 \times 10^{-4}$ \\
  & Weight decay         & $1 \times 10^{-4}$ \\
  & Batch size           & 256 \\
  & Hidden layers        & 256--256--256--128 \\
  & Activation           & SiLU \\
  & Max epochs           & 50 \\

\midrule

\multirow{7}{*}{SWAG Fine-tuning}
  & Optimizer            & SGD \\
  & Initial learning rate & $1 \times 10^{-2}$ \\
  & Momentum             & 0.9 \\
  & Weight decay         & $1 \times 10^{-4}$ \\
  & Batch size           & 256 \\
  & SWAG collection      & One model per full epoch \\
  & Max. collected models & 40 \\

\midrule

\multirow{3}{*}{\parbox{4.1cm}{SWAG LR Exploration}}
  & Learning-rate range  & Base LR to ($3\times$ base) LR \\
  & Cycle length         & 4 epochs \\
  & Schedule             & Triangular cyclic \\

\midrule

\multirow{3}{*}{\parbox{4.1cm}{LR Scheduler\\(all autoencoders)}}
  & Patience             & 8 epochs \\
  & Reduction factor     & 0.5 \\
  & Minimum LR           & $1 \times 10^{-5}$ \\

\bottomrule
\end{tabular}

\end{table}
\newpage
\section{Soil and Location Properties}

Soil properties for each location were obtained from SSURGO databases using APSIM-based soil profile extraction \cite{usda1999united}, providing depth-resolved hydraulic and chemical parameters. To capture a representative range of soil conditions, candidate locations were embedded in a standardized soil-property space and a farthest-point sampling strategy was applied to select six sites that maximize pairwise dissimilarity across key soil attributes.
\begin{center}
\begin{minipage}{0.95\textwidth}

\footnotesize
\setlength{\tabcolsep}{3pt}
\renewcommand{\arraystretch}{1.1}

\captionof{table}{Soil and location properties}
\label{tab:soil_location_properties}

\centering
\begin{tabular}{lrrrrrrrrrrl}
\toprule
County & Lon & Lat & DUL & LL15 & Carbon &  FInert & CN2 & SAT & BD & Soil \\
\midrule
Randolph  & -90.0600 & 38.0567 & 0.3372 & 0.1086 & 0.0118  & 2 & 73 & 0.4751 & 1.3884 & loam \\
Mason     & -89.7201 & 40.3537 & 0.0967 & 0.0406 & 0.0100  & 2 & 60 & 0.4257 & 1.5200 & sandy \\
Poweshiek & -92.2234 & 41.5487 & 0.3022 & 0.1718 & 0.0100  & 2 & 92 & 0.4285 & 1.5117 & loam \\
Bremer    & -92.1385 & 42.8147 & 0.3134 & 0.1631 & 0.0293  & 1 & 99 & 0.5205 & 1.2681 & silt loam \\
Logan     & -89.4897 & 39.9315 & 0.3661 & 0.1604 & 0.0118  & 2 & 74 & 0.4964 & 1.3318 & silt loam \\
Osceola   & -95.8179 & 43.3630 & 0.3977 & 0.2216 & 0.0287  & 1 & 99 & 0.5761 & 1.1208 & clay loam \\
\bottomrule
\end{tabular}

\end{minipage}
\end{center}

\section{Yield Distributions Across Projection Years and Emissions Scenarios}

Each figure compares the same 100,000 maize trait configurations under the
2020 historical baseline and under future projections from four climate models
across six Iowa--Illinois locations and four future projection periods. The
distributions are shown after coefficient-of-variation filtering ($CV<0.4$),
emphasizing predictions for which the emulator uncertainty is small relative to
the predicted yield.

Under SSP245, the 2040 distributions remain relatively close to the 2020
  control at most locations, although the climate models already produce
  different shifts in yield. By 2060, these differences become more visible:
  some locations retain substantial overlap with the control, whereas others
  shift toward lower yields or show greater model-to-model separation. The
  downward shift is generally stronger in 2080 and 2100, particularly at
  Randolph, Poweshiek, and Mason, while Logan and Bremer retain comparatively
  higher yields under several models. The changing widths of the violins also
  show that the projected climate affects not only the median yield but also
  the spread of outcomes across trait configurations. Thus, under SSP245,
  future yield loss develops progressively, but its magnitude depends strongly
  on location and climate model.

\begin{figure}[H]
    \makebox[\textwidth][c]{%
        \includegraphics[width=1.08\textwidth]{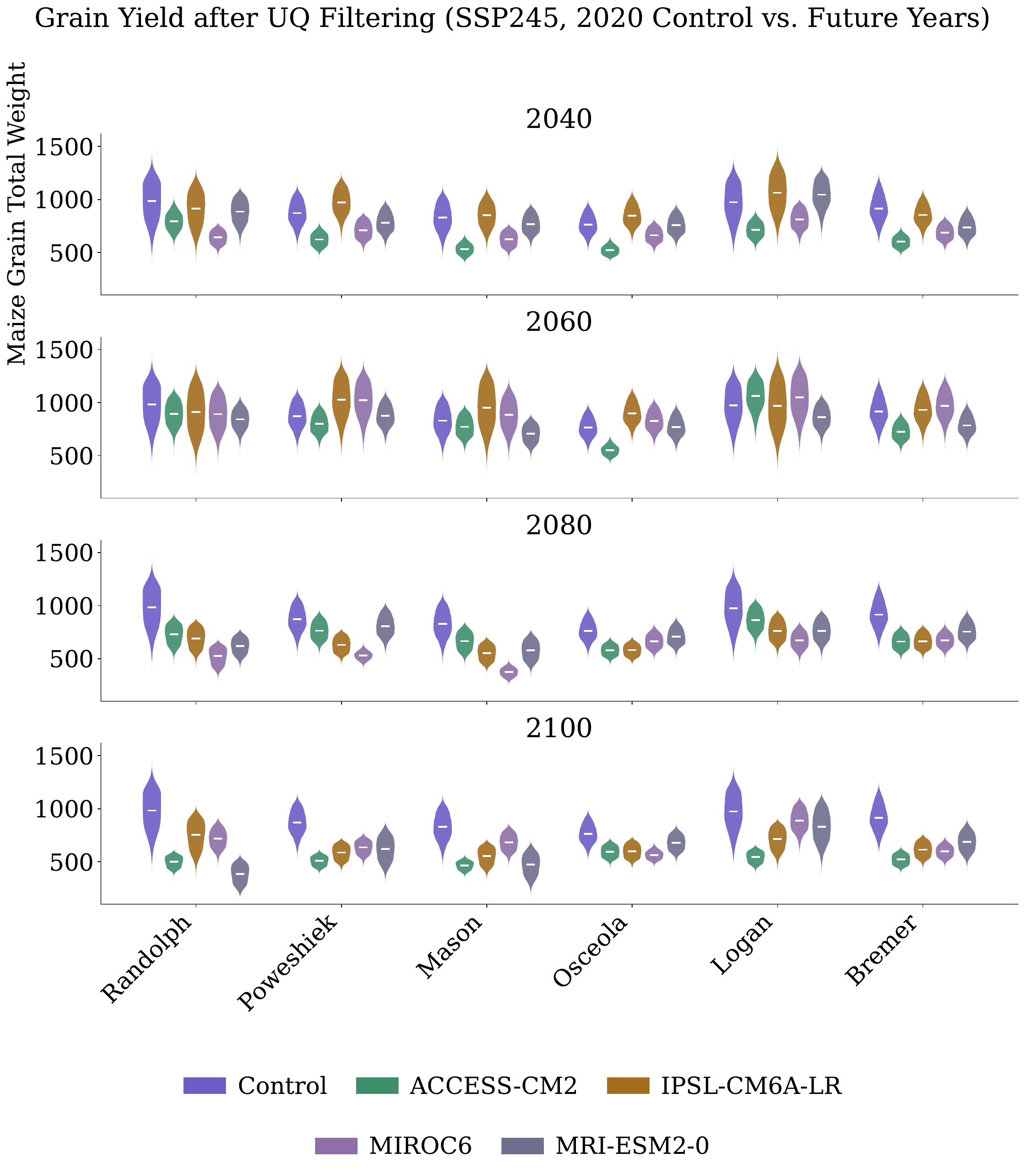}
    }
    \captionsetup{width=0.95\textwidth}
    \caption{Yield distributions under SSP245 across 2040, 2060, 2080, and 2100. Each panel compares the 2020 baseline with four climate models across six Iowa--Illinois locations after coefficient-of-variation filtering.}
    \label{fig:violin_SSP245_4years}
\end{figure}

\clearpage

Under SSP585, the stronger forcing pathway produces larger and more
location-dependent changes in yield level and distributional spread. The
separation between future distributions and the 2020 baseline becomes more
pronounced for several location--model combinations, while others retain
substantial overlap with the baseline. Comparing the four panels shows whether
the response emerges progressively or is concentrated near the end of the
century. Differences among the four climate models further demonstrate that
the magnitude of future yield change depends not only on the emissions
scenario, but also on the particular climate trajectory and local soil
environment.

\begin{figure}[H]
    \makebox[\textwidth][c]{%
        \includegraphics[width=0.90\textwidth]{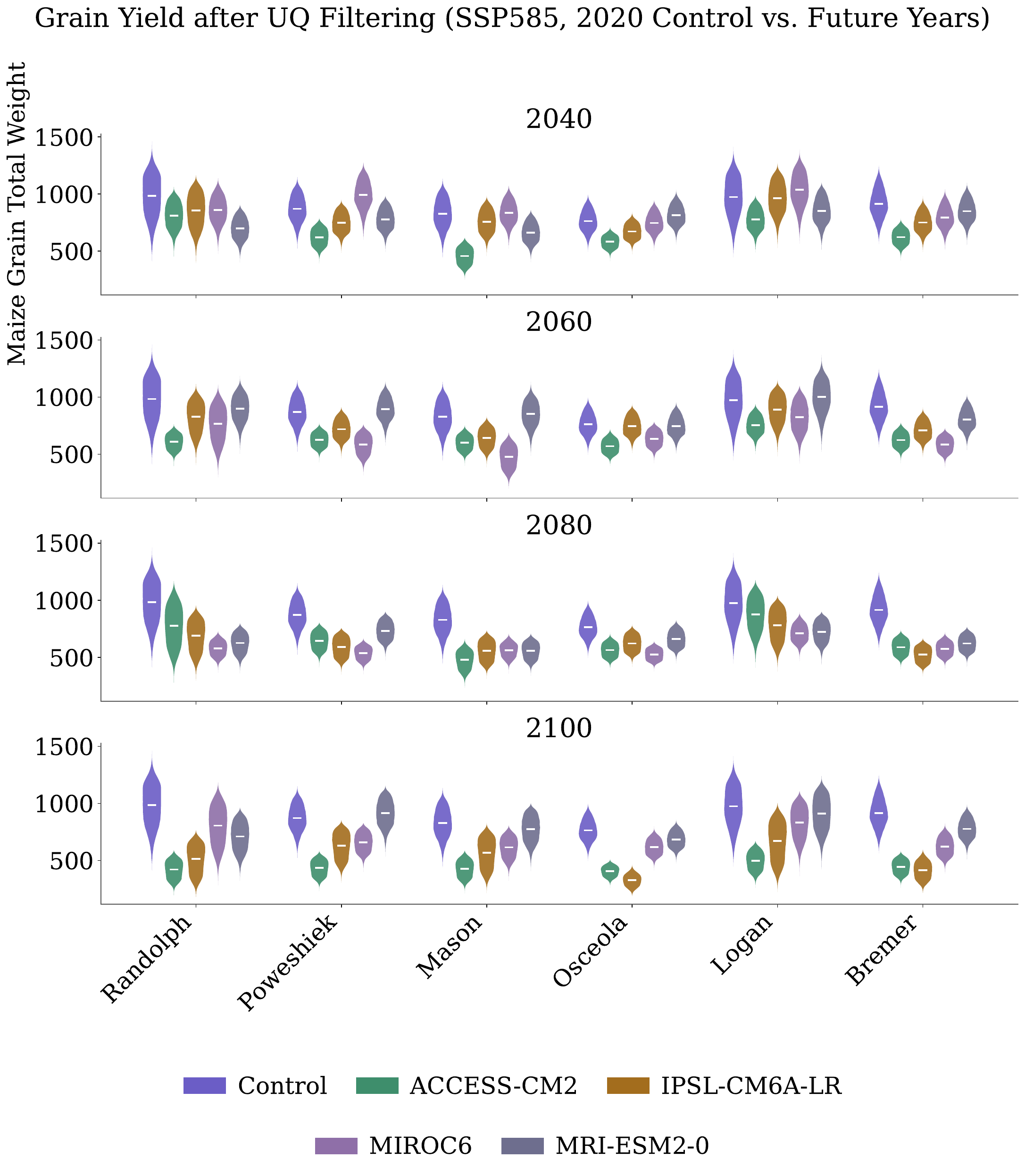}
    }
    \captionsetup{width=0.95\textwidth}
    \caption{Yield distributions under SSP585 across 2040, 2060, 2080, and 2100. Each panel compares the 2020 baseline with four climate-model projections across six Iowa--Illinois locations after coefficient-of-variation filtering, showing location-, model-, and year-dependent changes in yield magnitude and spread.}
    \label{fig:violin_SSP585_4years}
\end{figure}

\clearpage
\section{Trait Importance Analysis}
The heatmaps summarize permutation importance for the five grouped trait
categories across six locations and four projection years. Each cell reports
the rank followed by the percentage contribution, so darker cells indicate
traits that explain more of the variation in predicted yield. The values are
averaged across the four climate models within each emissions scenario.

The SSP245 panels show from 2040 to 2100, RUE remains the leading contributor in both scenarios and
at every location, but its relative importance changes over time. Under
SSP245, the strongest shift occurs around 2060, when RUE decreases to
approximately $55$--$69\%$ and MGPC becomes more influential, reaching
approximately $28$--$34\%$ at several locations. By 2080 and 2100, the
importance of RUE increases again to roughly $72$--$84\%$, while the
contribution of MGPC generally declines.

\begin{figure}[H]
    \makebox[\textwidth][c]{%
        \includegraphics[width=1.14\textwidth]{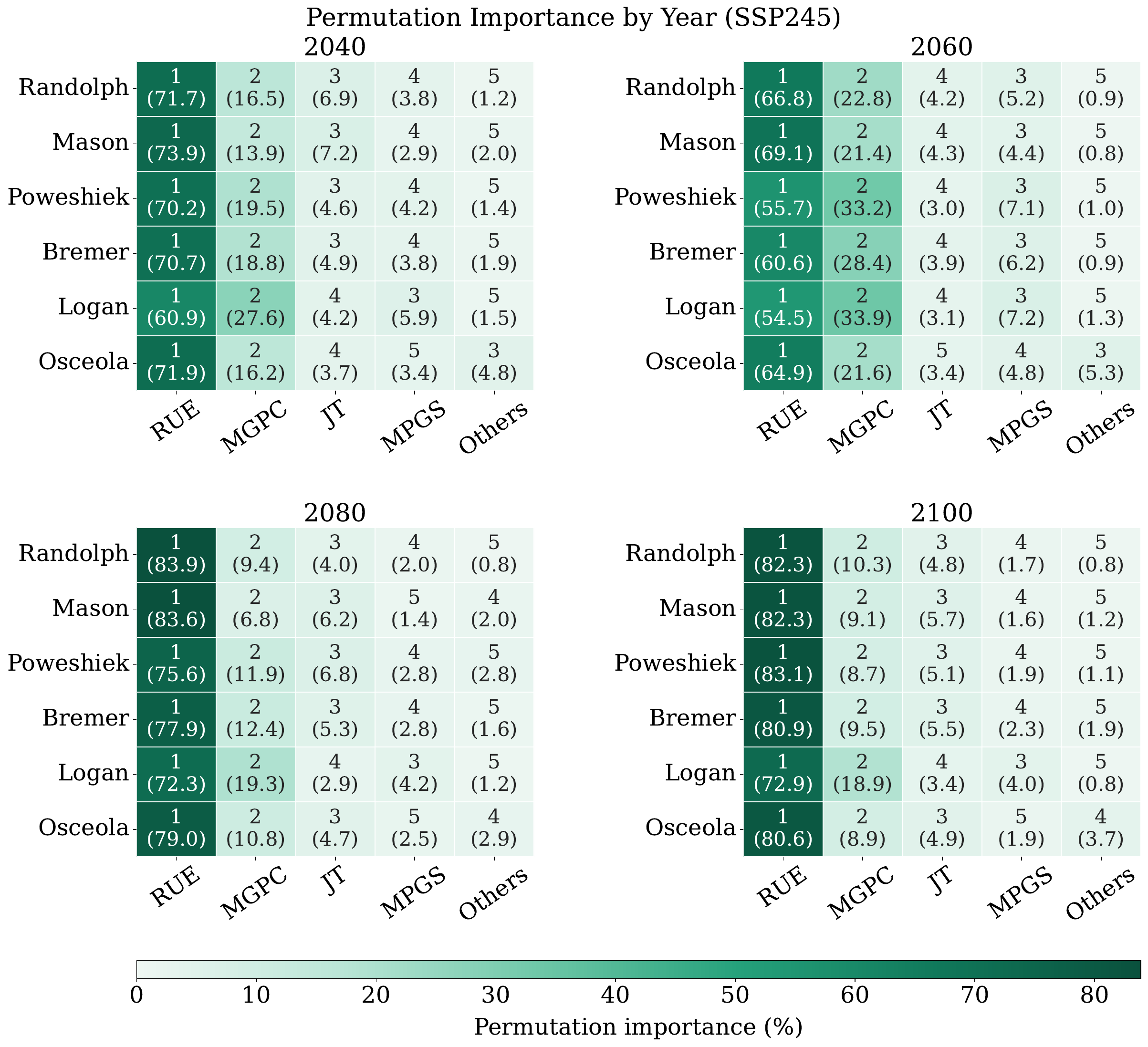}
    }
    \captionsetup{width=1\textwidth}
    \caption{Permutation importance of grouped genetic traits under SSP245 across the projection years. The x-axis lists the grouped genotype traits and the y-axis lists the locations; each panel shows location-specific importance averaged across the four climate models. Cell labels report rank followed by importance (\%). RUE denotes radiation use efficiency, MGPC denotes maximum grains per cob, JT denotes Juvenile+Target, and MPGS denotes maximum potential grain size. ``Others'' represents the remaining grouped traits not shown individually.}
    \label{fig:permutation_SSP245}
\end{figure}

\clearpage

  SSP585 produces a faster transition toward RUE dominance during the middle
  projection period, although the two scenarios become more similar by 2100.
  The main persistent exception is Logan, where MGPC remains relatively
  important throughout the century. JT, MPGS, and the ``Others'' category remain
  secondary in both scenarios, generally contributing only small shares and
  showing location-specific changes rather than a consistent long-term trend.
  Overall, the figures suggest that the scenarios mainly differ in the timing
  and strength of the temporary reweighting between RUE and MGPC, while the
  identity of the dominant trait remains unchanged.

\begin{figure}[H]
    \makebox[\textwidth][c]{%
        \includegraphics[width=1.2\textwidth]{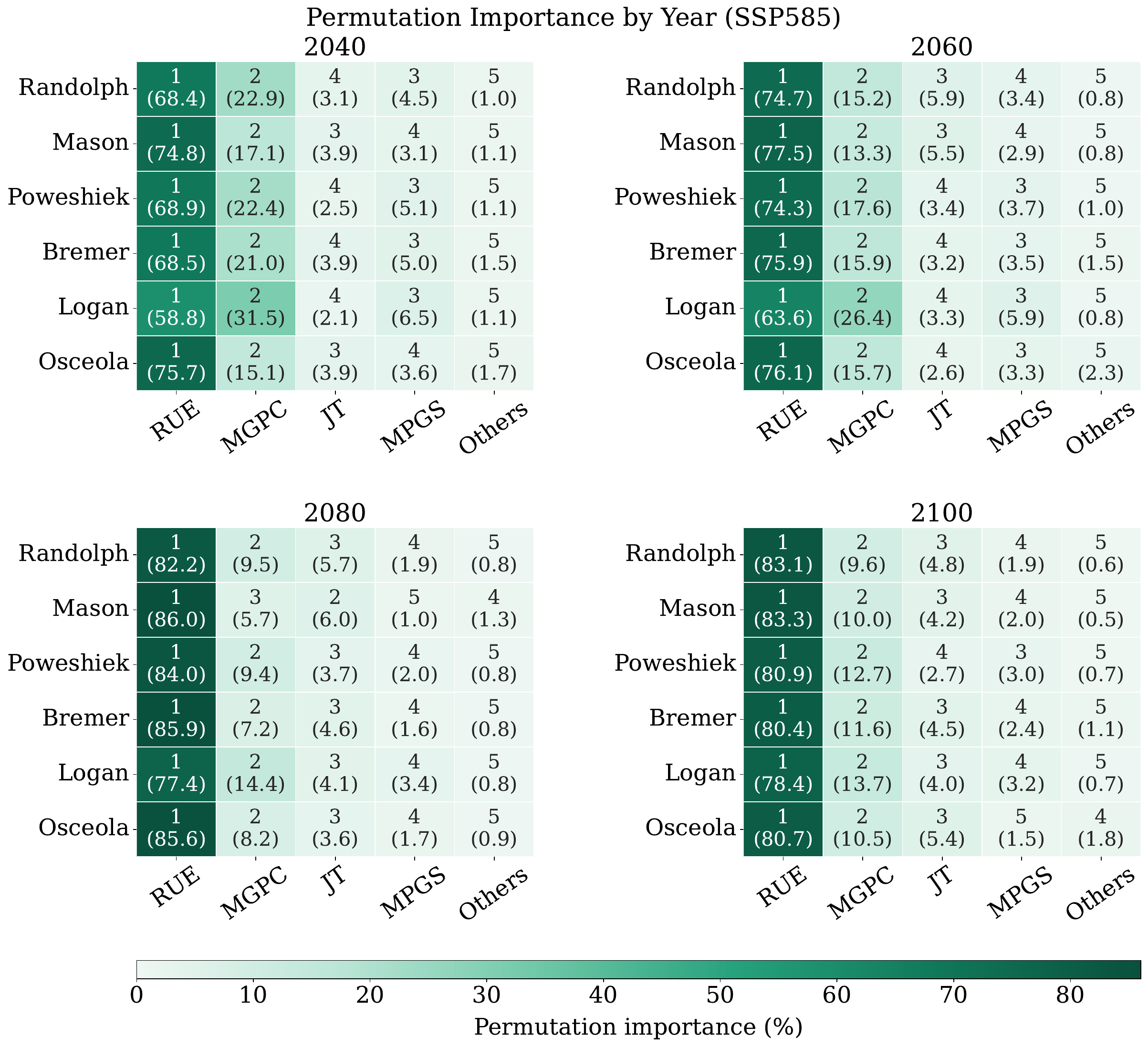}
    }
    \captionsetup{width=1\textwidth}
    \caption{Permutation importance of grouped genetic traits under SSP585 across the projection years. The x-axis lists the grouped genotype traits and the y-axis lists the locations; each panel shows location-specific importance averaged across the four climate models. Cell labels report rank followed by importance (\%). RUE denotes radiation use efficiency, MGPC denotes maximum grains per cob, JT denotes Juvenile+Target, and MPGS denotes maximum potential grain size. ``Others'' represents the remaining grouped traits not shown individually.}
    \label{fig:permutation_SSP585}
\end{figure}

\clearpage
\section{Resilient Trait Distributions Across Projection Years}

The resilient configurations identified for 2100 show a pronounced concentration
of radiation use efficiency (RUE) near the upper end of its sampled range.
Several yield-related traits also
shift toward higher values: Juvenile+Target is concentrated near the upper end
of its range, MaximumGrainsPerCob is concentrated toward high values, and
MaximumPotentialGrainSize is enriched toward the upper part of its range.
Final N concentration shifts toward lower values, whereas ShootLag, ShootRate,
FloweringToGrainFilling+Target, GrainFilling+Target, the two temperature factors,
and PotentialExtinctionCoeff retain broader distributions. Thus, the 2100
resilient set combines a common high-RUE profile with selective shifts in
developmental, reproductive, and nitrogen-related traits rather than maximizing
all traits simultaneously.

\begin{figure}[H]
    \centering
    \includegraphics[width=0.78\textwidth]{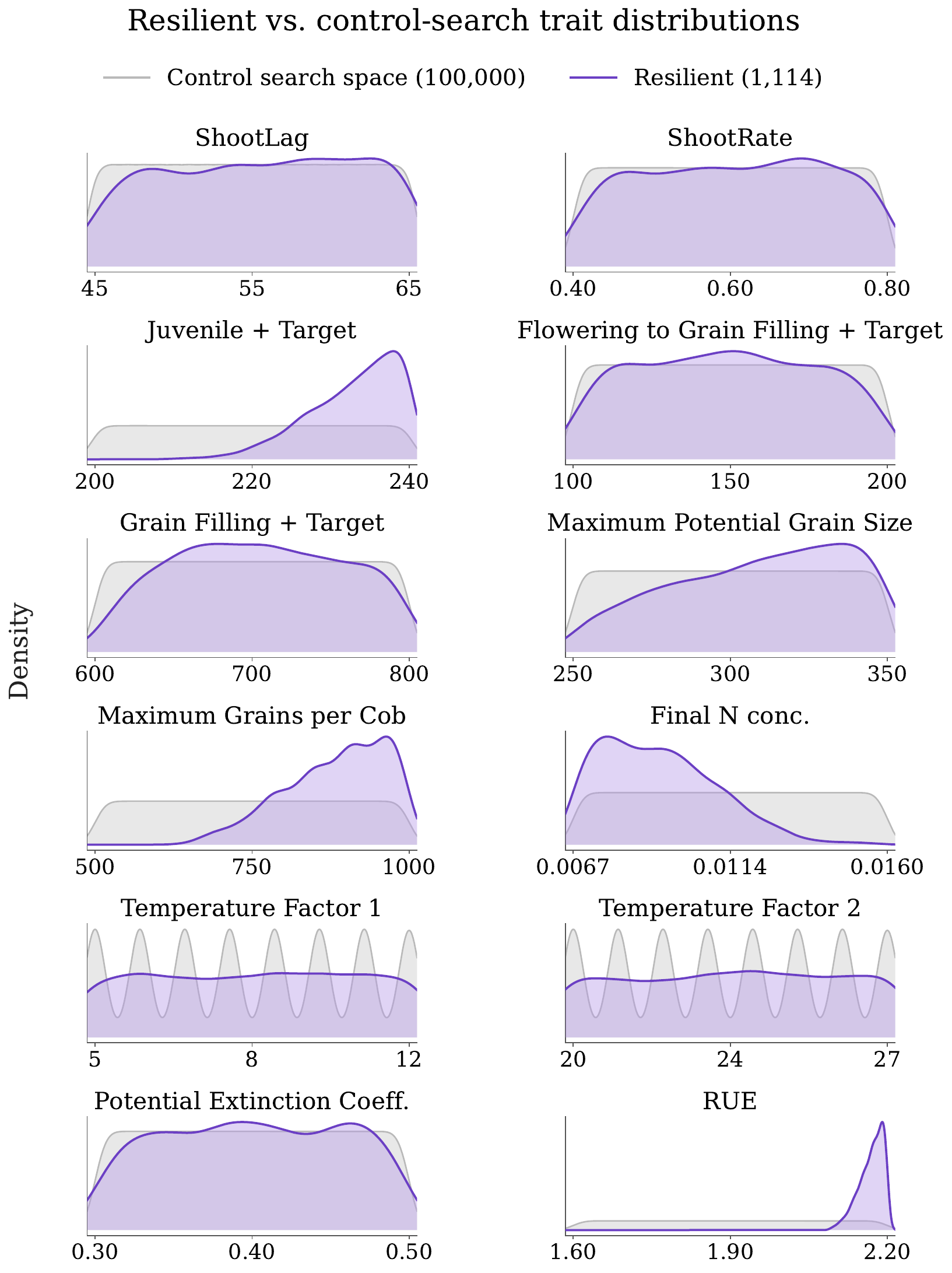}
    \captionsetup{width=1\textwidth}
    \caption{Distributions of the 12 sampled traits for the resilient configurations identified for 2100 compared with the full 100,000-configuration control search space. The resilient set contains 1,114 configurations.}
    \label{fig:resilient_distribution_2100}
\end{figure}

\clearpage

When resilience is required across the four future projection years (2040, 2060, 2080,
and 2100), excluding the 2020 historical baseline, the resilient set is smaller, comprising 246 configurations compared
with 1,114 configurations identified for 2100 alone. The high-RUE concentration
remains the most consistent feature, and the upward shifts in Juvenile+Target,
MaximumPotentialGrainSize, and MaximumGrainsPerCob are also retained. However,
the four-year set shows additional year-robust structure in traits such as
ShootLag, ShootRate, and the temperature factors, while
FloweringToGrainFilling+Target and GrainFilling+Target remain more broadly
distributed. These results indicate that requiring resilience across multiple
future years narrows the feasible trait space while preserving a common high-RUE
background and allowing flexibility in several secondary traits.

\begin{figure}[H]
    \centering
    \includegraphics[width=0.78\textwidth]{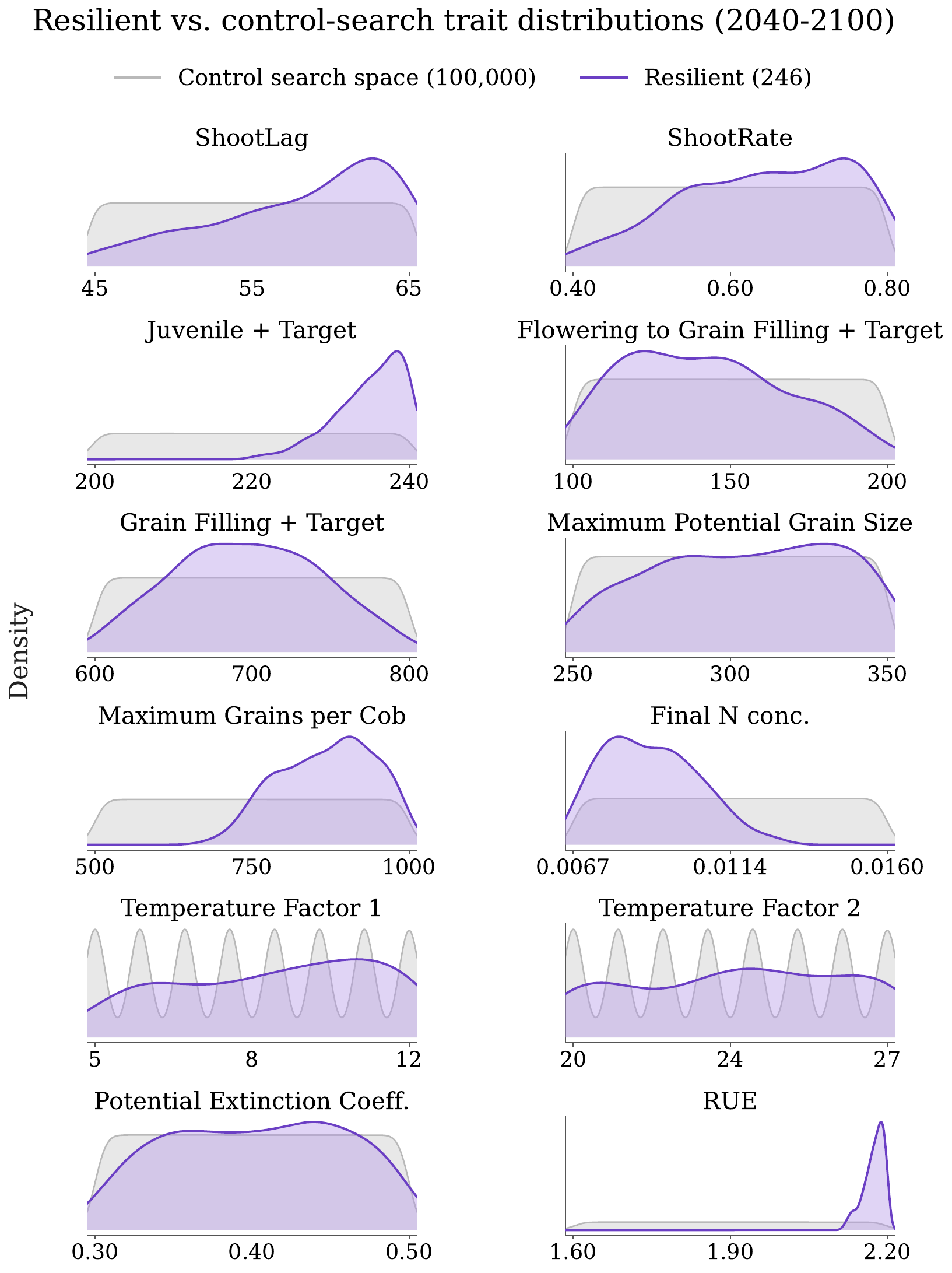}
    \captionsetup{width=1\textwidth}
    \caption{Distributions of the 12 sampled traits for configurations resilient across the four projection years (2040, 2060, 2080, and 2100) compared with the full 100,000-configuration control search space. The multi-year resilient set contains 246 configurations.}
    \label{fig:resilient_distribution_4y}
\end{figure}

\clearpage
\section{Correlation Structure of the 2100-Resilient Trait Set}

To examine how the selected traits vary together, we calculated pairwise
Pearson correlations for the 1,114 configurations resilient in 2100. Most
off-diagonal correlations are close to zero, indicating that resilience is
not defined by one broad set of strongly coupled traits. Instead, the
resilient population retains largely independent variation across many
traits, with a few moderate relationships: Juvenile+Target is positively
associated with Final N concentration ($r=0.27$), while MaximumPotentialGrainSize
and MaximumGrainsPerCob ($r=-0.29$), and Juvenile+Target and RUE ($r=-0.28$),
show moderate negative associations. These relationships suggest that the
resilient set reflects a combination of specific trait trade-offs rather than
a uniform shift of all traits in the same direction.

\begin{figure}[H]
    \makebox[\textwidth][c]{%
        \includegraphics[width=1.2\textwidth]{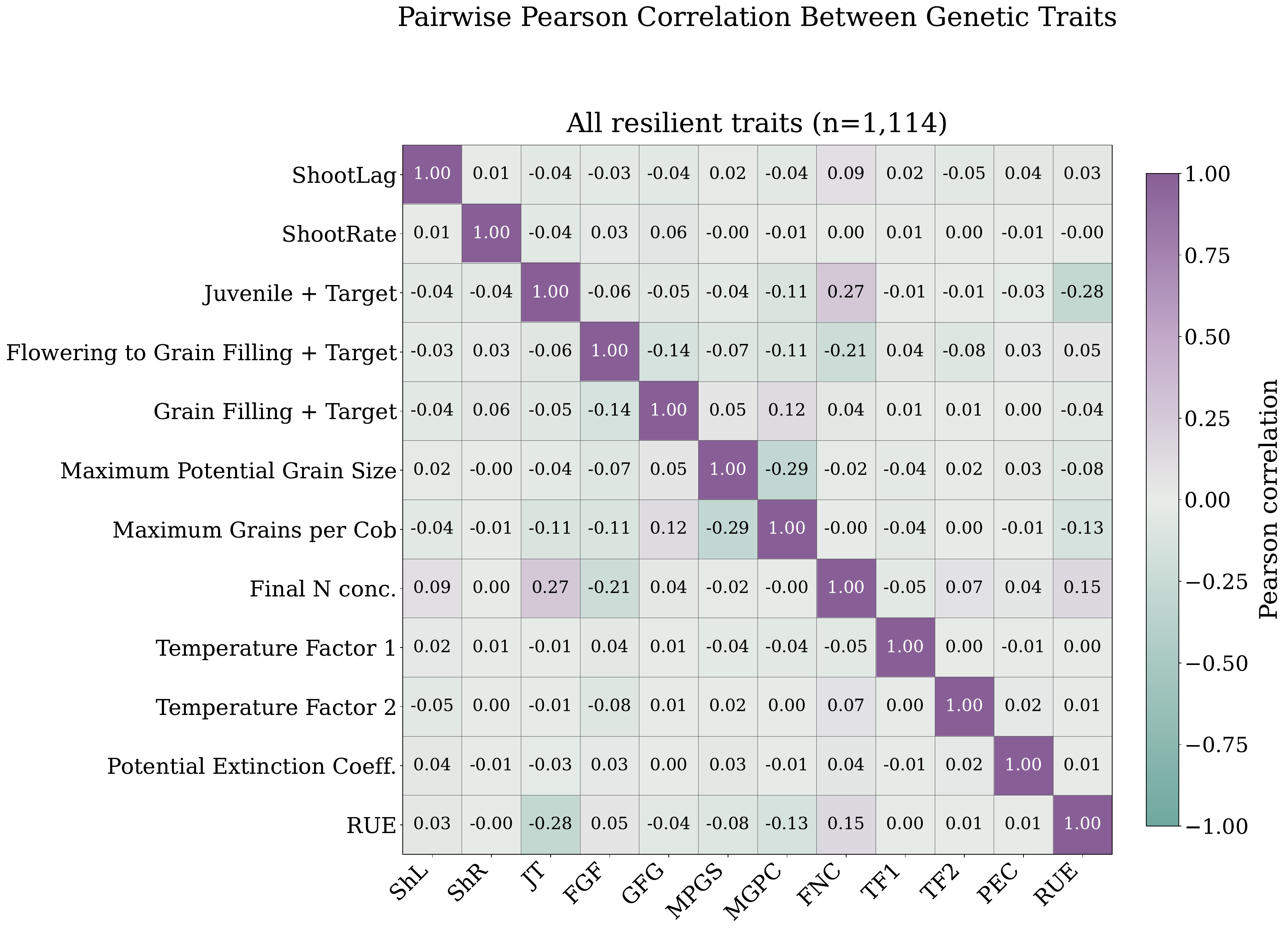}
    }
    \captionsetup{width=1\textwidth}
    \caption{Pairwise Pearson correlations among the 12 genetic traits for the
    1,114 configurations resilient in 2100. The diagonal values are the
    self-correlations ($r=1$), while the off-diagonal values show how pairs of
    traits vary across the resilient population. Most pairwise relationships
    are weak, with only a small number of moderate positive or negative
    associations.}
    \label{fig:resilient_trait_correlation_2100}
\end{figure}

\clearpage
\section{Multivariate Structure of Resilient Trait Configurations}

To compare the overall genetic-trait composition of the resilient sets, we
performed principal component analysis (PCA) using the 12 sampled genetic
traits. PCA was performed separately for each comparison shown in the figure,
using the control population together with the corresponding resilient set.
In each analysis, the resilient population defines the standardized trait space
and the PCA axes, while the control population is projected into that
resilient-defined space for reference. The resulting projections provide a
compact view of the internal multivariate structure of each resilient set and
its relationship to the broader control population.

The PCA view makes the increasing selectivity of resilience visible. The
2100-resilient configurations occupy a relatively peripheral tail of the
control distribution rather than its central mass. When resilience is required
across all four projection years, this tail becomes smaller, narrower, and more
organized along the dominant direction of trait variation. Thus, the four-year
criterion does more than reduce the number of configurations: it concentrates
resilience into a tighter region of the joint trait space, consistent with the
shared high-RUE and favorable developmental and reproductive-trait profile
identified in the univariate distributions. This concentration is also
reflected in the modest increase in the variance carried by the first two
components: PC1 and PC2 explain 11.96\% and 11.16\% for the 2100-resilient
set, compared with 14.89\% and 12.86\% for the four-year set.

\begin{figure}[H]
    \centering
    \includegraphics[width=1\textwidth]{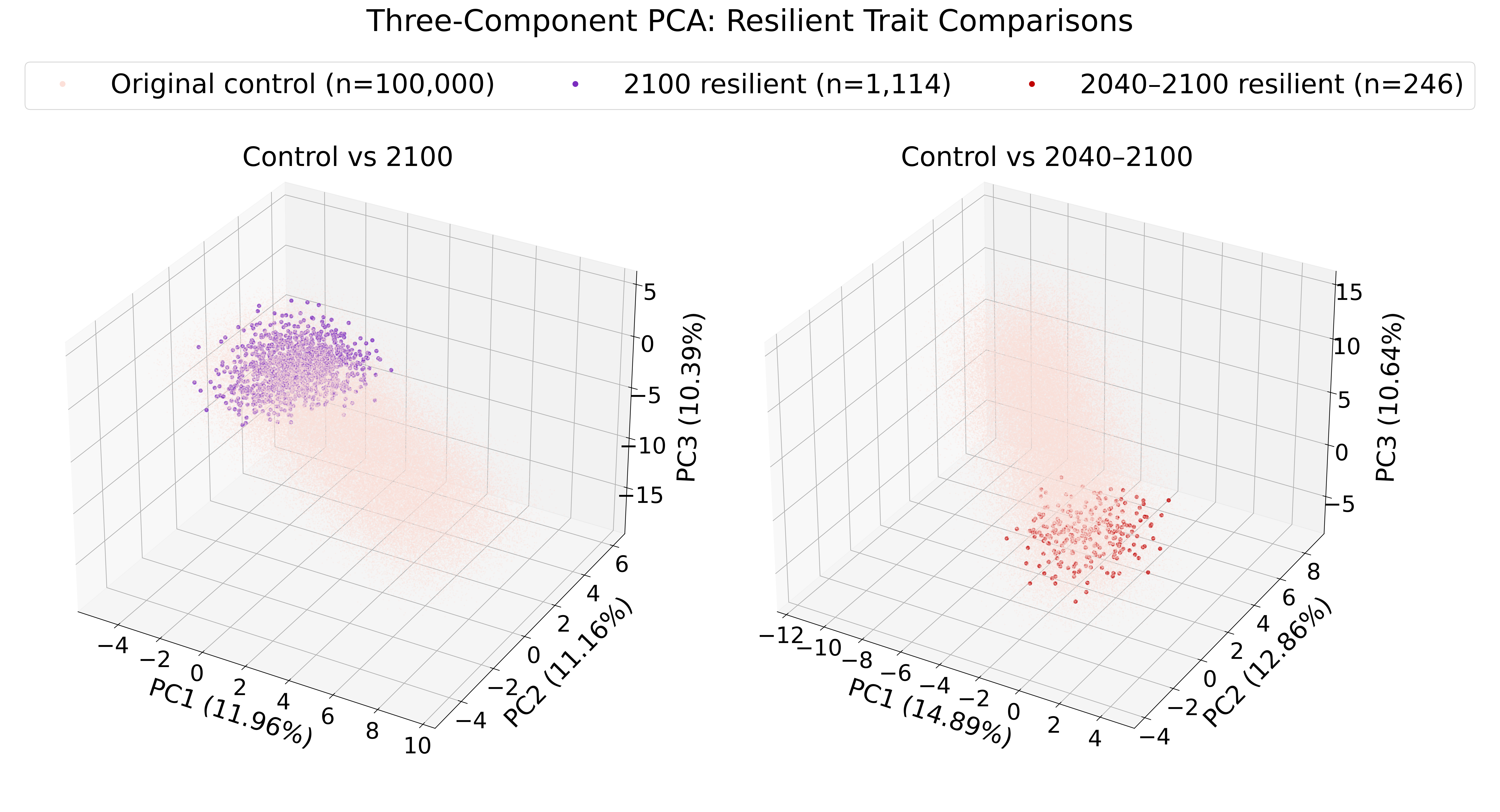}
    \captionsetup{width=1\textwidth}
    \caption{Three-component PCA comparisons of the 12 genetic traits. The green
    points show the original control population of 100,000 configurations.
    Purple points show the 1,114 configurations resilient in 2100, while red
    points show the 246 configurations resilient across all four projection
    years (2040, 2060, 2080, and 2100). The PCA projections show that resilient
    configurations remain embedded within the feasible control population while
    occupying a more structured multivariate region. PCA is fitted to each
    resilient population, and the control population is projected into the
    corresponding resilient-defined space. The first three components explain
    11.96\%, 11.16\%, and 10.39\% of the variance for the 2100 resilient set,
    and 14.89\%, 12.86\%, and 10.64\% for the 2040--2100 resilient set,
    respectively.}
    \label{fig:pca_resilient_control_comparison}
\end{figure}

\clearpage
\section{Direct APSIM Verification of the Four-Year Resilient Configurations}

The emulator was used to identify resilient trait configurations because it
allows the large trait--environment space to be explored at a fraction of the
cost of direct APSIM simulation. Nevertheless, configurations retained by this
screening step should be checked against the mechanistic simulator itself.
This verification is important because approximation errors introduced during
emulator training, or arising from the finite and necessarily coarse coverage
of the input space, could propagate into the final resilient set. In particular,
future weather conditions and the high-performing trait combinations retained
as resilient configurations occupy relatively sparse upper-tail regions of the
broader weather and response distributions, where emulator error may be less
uniform. Because selection is based on rank ordering, substantial disagreement
between emulator and APSIM rankings could also alter which configurations are
classified as resilient. Direct comparison with APSIM therefore tests whether
tail-region errors materially affect yield predictions or the resulting resilient
selection, while confirming agreement with the mechanistic simulator that
generated the training targets.

We performed this verification for the 246 configurations that remained in
the intersection across the four future projection years (2040, 2060, 2080, and
2100). Each configuration was evaluated at six locations under 32 future
weather combinations, comprising four climate models, two emissions scenarios,
and four future projection years. The 2020 control weather files were not
included in this verification. This resulted in 47,232 APSIM simulations. APSIM
grain yield was extracted from the completed report file. We retained the
signed difference, defined as APSIM yield minus emulator yield, only to quantify
directional bias; absolute yield errors were used to quantify prediction-error
magnitude through MAE and RMSE.

For an additional trait-level summary, APSIM yields were first averaged across
the 192 verified environments for each of the 246 finalist configurations and
then compared with the corresponding emulator yield summary. We calculated one
signed difference per finalist as APSIM mean yield minus emulator mean yield.
The resulting differences were positive for the finalist set, with a mean
difference of 50.0~g~m$^{-2}$ and a median difference of 50.3~g~m$^{-2}$;
positive values indicate that APSIM predicted a higher mean yield than the
emulator. Thus, the box plot shows the distribution of 246 trait-level
differences, rather than 192 separate environment-level distributions.

\begin{figure}[H]
    \centering
    \includegraphics[width=0.95\textwidth]{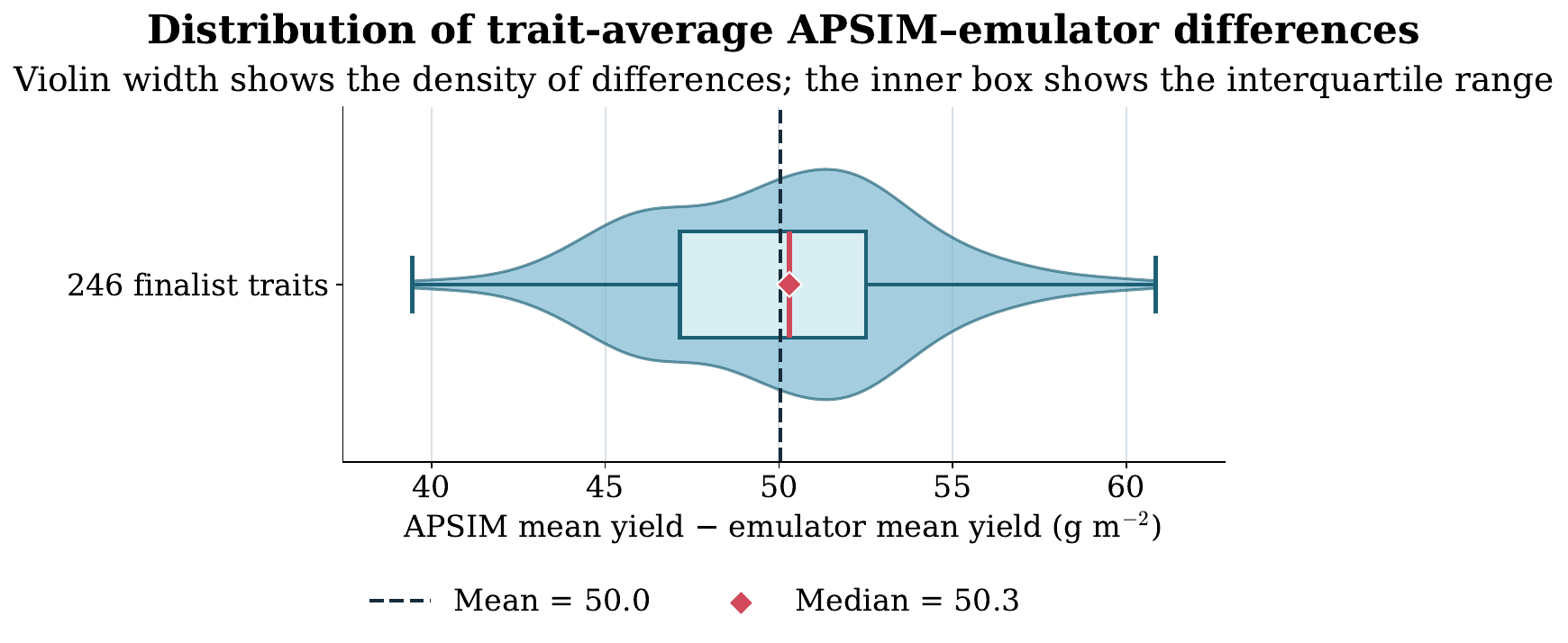}
    \captionsetup{width=1\textwidth}
    \caption{Distribution of trait-average APSIM--emulator yield differences for the 246 four-year resilient configurations. Differences are calculated as APSIM mean yield minus emulator mean yield after averaging across the verified future environments for each configuration. The violin width shows the distribution density, the inner box shows the interquartile range, and the markers indicate the mean and median. Positive values indicate that APSIM predicted higher mean yield than the emulator.}
    \label{fig:apsim_emulator_trait_average_difference}
\end{figure}

\clearpage

To determine whether emulator errors were hidden by averaging across
environments, we compared APSIM and emulator predictions for every matched
finalist--environment case. This case-level comparison preserves the
environmental structure of the verification experiment and shows whether
errors are concentrated under particular climate models, locations, years, or
emissions scenarios. It therefore provides a direct test of both numerical
agreement with APSIM and the possibility that environment-specific errors could
affect the resilient ranking.

The exact case-level comparison resulted in an overall normalized RMSE of 12.3\%,
with variation across climate-model--scenario combinations, locations,
projection years, and emissions scenarios. The largest normalized RMSE values
occurred for the MRI-ESM2-0--SSP245 combination (21.6\%), Poweshiek County (18.7\%),
and the 2100 projection year (19.4\%), whereas other strata showed lower
errors. These results indicate that emulator--APSIM discrepancies are
environment-dependent rather than uniform across the verification domain. The
overall error indicates that the emulator retains useful quantitative agreement
with APSIM for evaluating the finalist set, while the larger errors in selected
climate-model, location, and late-century strata identify conditions where the
emulator is less precise. This pattern is consistent with the greater difficulty
of representing relatively sparse or more extreme regions of the weather and
response distributions, while the overall agreement supports the use of the
emulator for cross-environment resilient selection.

\begin{figure}[H]
    \centering
    \includegraphics[width=1\textwidth]{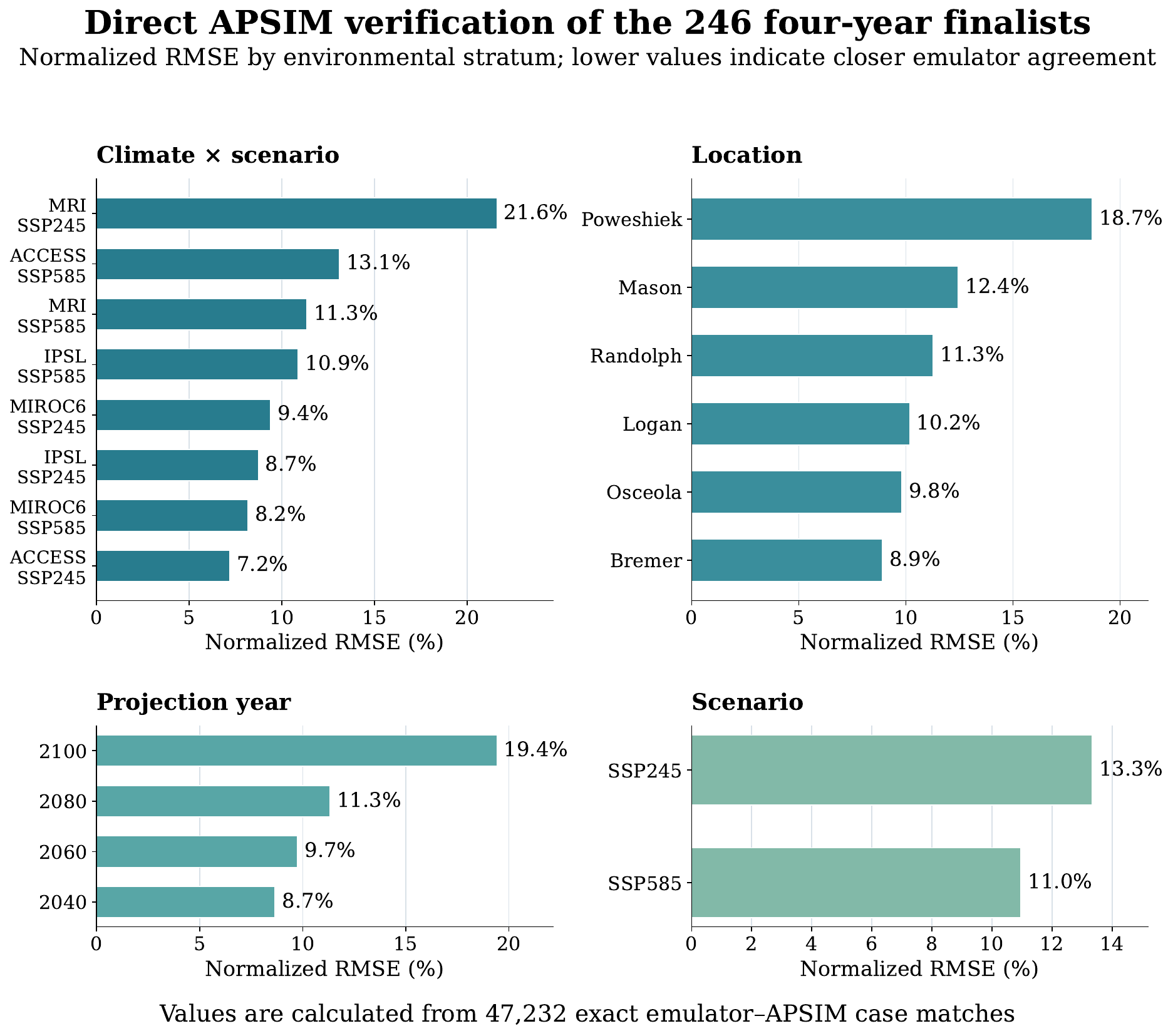}
    \captionsetup{width=1\textwidth}
    \caption{Stratified normalized root-mean-square error (NRMSE) for the exact emulator--APSIM comparison of the 246 four-year resilient configurations. Values are calculated from 47,232 matched configuration--environment cases and are shown by climate-model--scenario combination, location, projection year, and emissions scenario. Lower values indicate closer agreement between emulator and APSIM.}
    \label{fig:apsim_emulator_normalized_rmse_strata}
\end{figure}


Because the resilient configurations were selected through repeated ranking
across environments, agreement between emulator and APSIM rankings is an
important complement to yield-error metrics. Rank-order differences warrant
verification because a large discrepancy could alter the selected resilient
set, whereas small local changes among closely competing configurations are not
necessarily consequential when the screening retains a broad top-5,000 group
out of 100,000 configurations. We therefore calculated the absolute rank
distance between emulator and APSIM for every finalist in every environment.
Across the 47,232 individual finalist--environment comparisons (246 finalists
evaluated in each of 192 environments), the mean absolute rank distance was
48.0 ranks and the median was 37 ranks. These values indicate
moderate fine-scale disagreement rather than a wholesale reversal of the
ranking, and do not by themselves imply that the emulator failed to identify
high-performing configurations. The broad top-5\% screening criterion and the
cross-environment intersection therefore make the resilience conclusion robust
to such local rank changes. The stratified results further show that rank
distances vary across locations,
climate models, scenarios, and projection years. The scenario effect was small
(47.8 ranks for SSP245 versus 48.2 ranks for SSP585), whereas the location and
climate-model effects were larger, with the greatest mean distances for Osceola
County (59.8 ranks) and ACCESS-CM2 (54.4 ranks), and the smallest for Logan County
(38.7 ranks) and IPSL-CM6A-LR (42.8 ranks). Rank distance also increased modestly
toward the later projection years (47.2, 43.8, 50.2, and 50.8 ranks for 2040,
2060, 2080, and 2100, respectively), consistent with greater difficulty ranking
closely competing finalists under more extreme projected conditions. Overall,
the figure indicates environment-dependent fine-scale ranking differences
rather than a wholesale loss of agreement between the emulator and APSIM.

\begin{figure}[H]
    \centering
    \includegraphics[width=1\textwidth]{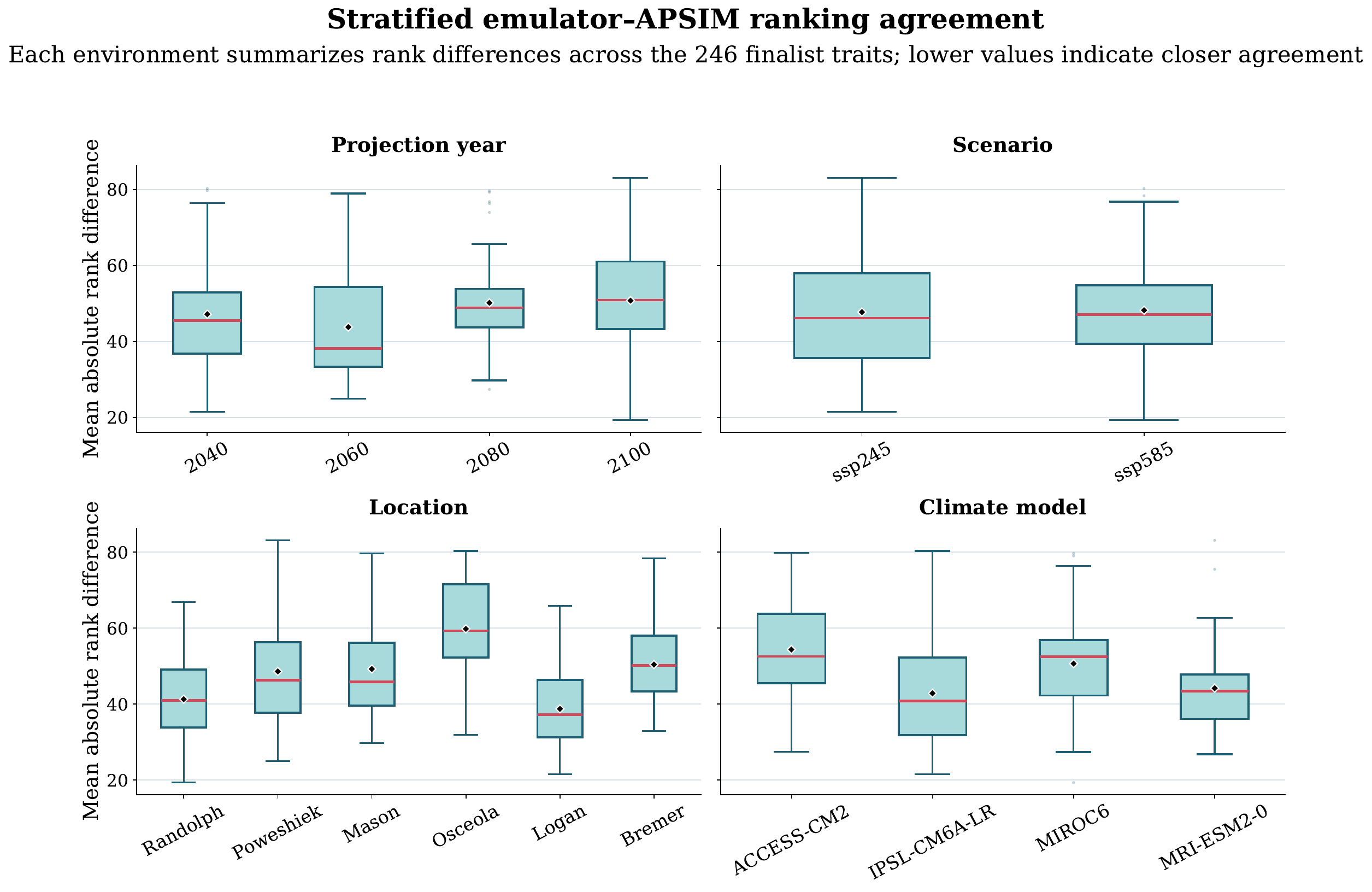}
    \captionsetup{width=1\textwidth}
    \caption{Stratified emulator--APSIM ranking disagreement for the 246 four-year resilient configurations. Each box summarizes the distribution of mean absolute rank differences across environments, where each environment summarizes the 246 finalist traits. Results are stratified by projection year, emissions scenario, location, and climate model. Lower values indicate closer agreement between emulator and APSIM; the horizontal line within each box indicates the median and the diamond indicates the mean.}
    \label{fig:apsim_emulator_stratified_rank_difference}
\end{figure}

\newpage

\section{Projected Climate Conditions}
We evaluate multiple climate models across six locations to characterize projected end-of-century conditions, with each model capturing distinct patterns in temperature and precipitation responses. While all models consistently indicate a warming trend, they differ in the magnitude of temperature increases and the intensity and frequency of precipitation events, providing a comprehensive representation of uncertainty and variability in future climate projections. Here, we present Randolph County under the SSP585 scenario as a representative example to illustrate how these patterns manifest across models.
\begin{figure}[H]
    \centering
    \includegraphics[width=1\textwidth]{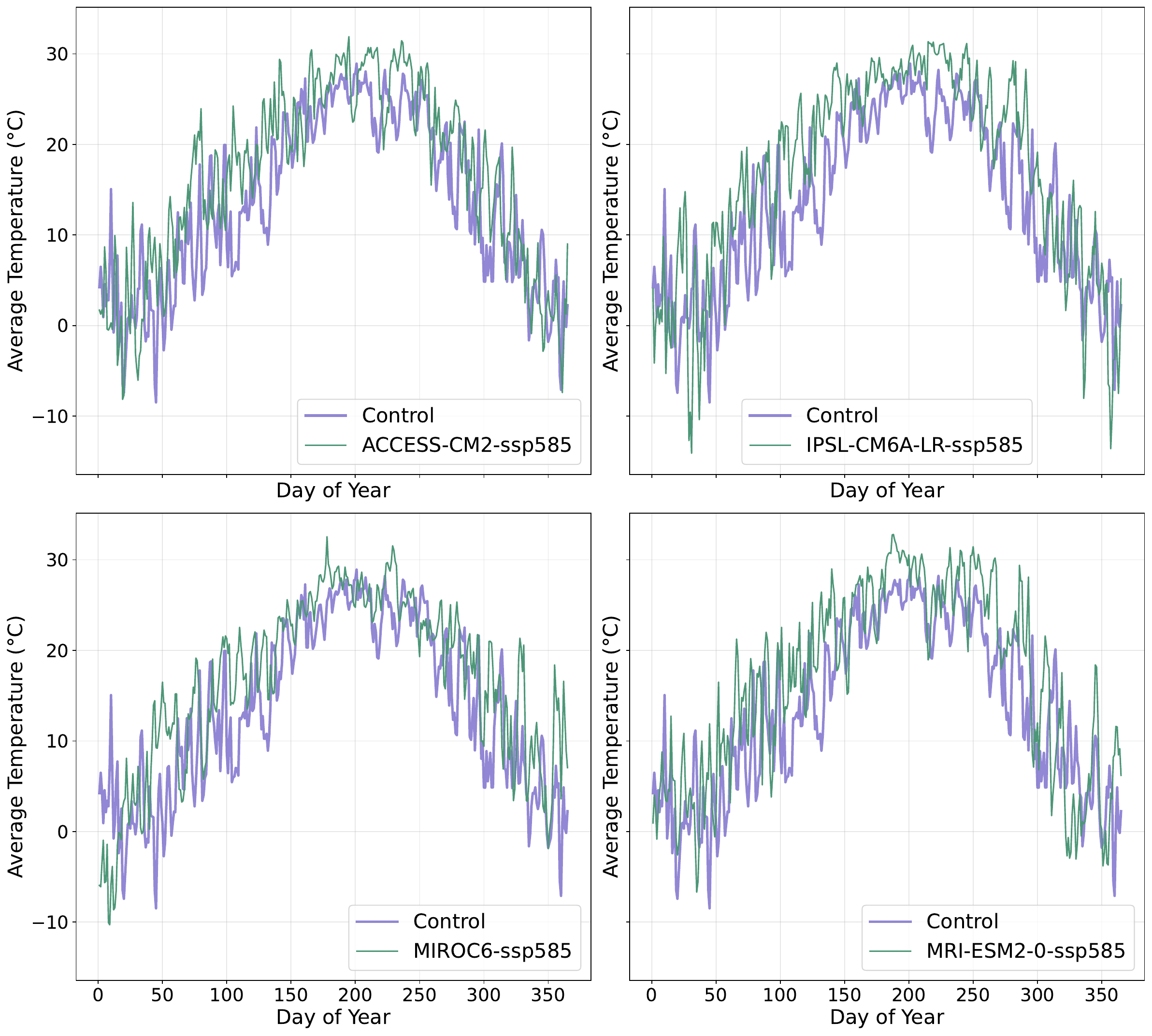}

    \caption{Comparison of daily average temperature between the 2020 historical baseline comparator and end-of-century projections for 2100 under the SSP585 scenario for Randolph County. Each panel represents a different climate model (ACCESS-CM2, IPSL-CM6A-LR, MIROC6, and MRI-ESM2-0), illustrating consistent warming trends across models, with higher temperatures and amplified seasonal peaks in the projected climate relative to the baseline weather.}
    \label{fig:belleville_weather_comparison}
\end{figure}

\begin{figure}[H]
    \centering
    \vspace{0.3cm}

    \includegraphics[width=1\textwidth]{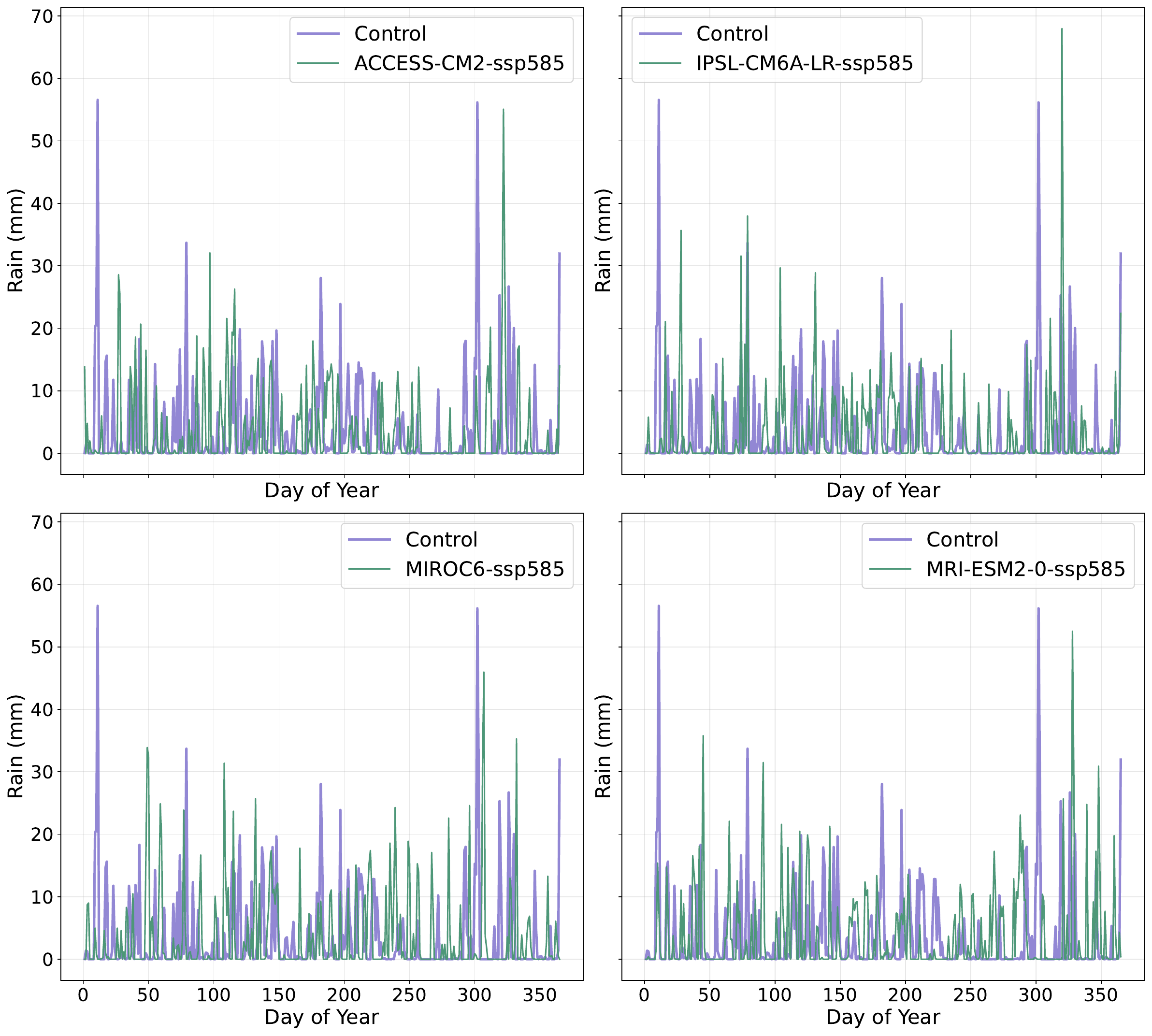}

    \caption{Comparison of daily precipitation between the 2020 historical baseline comparator and end-of-century projections for 2100 under the SSP585 scenario for Randolph County. Each panel represents a different climate model (ACCESS-CM2, IPSL-CM6A-LR, MIROC6, and MRI-ESM2-0), highlighting changes in rainfall patterns, including increased variability and more frequent high-intensity precipitation events in the projected climate relative to the baseline weather.}
    \label{fig:belleville_rain_comparison}
\end{figure}

\section{Coverage of Future Weather Within the Emulator Domain}
The downloaded archive of historical and projected weather sequences, spanning 1984--2100, contained 12,928 sequences, whereas the emulator training and subsequent exploration were conducted at a scale of approximately 2 million weather realizations. Using only the downloaded sequences would therefore provide a relatively small and discrete sample of the possible combinations of seasonal heat, rainfall, dry spells, and extreme events, increasing the risk of overfitting to particular weather realizations and limiting coverage of plausible weather variability. We addressed this limitation by generating synthetic weather sequences through convex sampling in the learned weather representation. This procedure increased the number of plausible weather realizations available to the emulator while remaining connected to the empirical weather archive.

Because the archived historical and future weather sequences provide the empirical reference for the analysis, it was important that the synthetic weather remain consistent with, and cover, the weather conditions represented in that archive. We therefore performed a validation after synthetic generation to confirm that the expanded synthetic dataset remained grounded in the archived weather data and that the future cases used for resilient-trait selection were represented within the same broader weather domain. For each weather sequence, we calculated four crop-relevant summary statistics: seasonal growing degree days using a 10\textdegree C base temperature, seasonal rainfall total, the number of days with maximum temperature above 35\textdegree C, and the 95th percentile of daily rainfall. We compared these statistics for the original weather archive, the 1.8 million convexly sampled synthetic sequences, and the 48 future location--climate-model--scenario files used as the year-2100 cases using empirical cumulative distribution functions (ECDFs). The ECDFs provide a direct comparison of the relative position and overlap of these three weather populations. The future cases occupy a different part of the broader weather distribution, but remain within the overall encoder-training range for all four statistics. The comparison therefore evaluates whether synthetic sampling preserves and broadens the finite set of weather realizations used in model development, rather than treating the future cases as an unseen historical--future category.

\begin{figure}[H]
    \centering
    \includegraphics[width=1\textwidth]{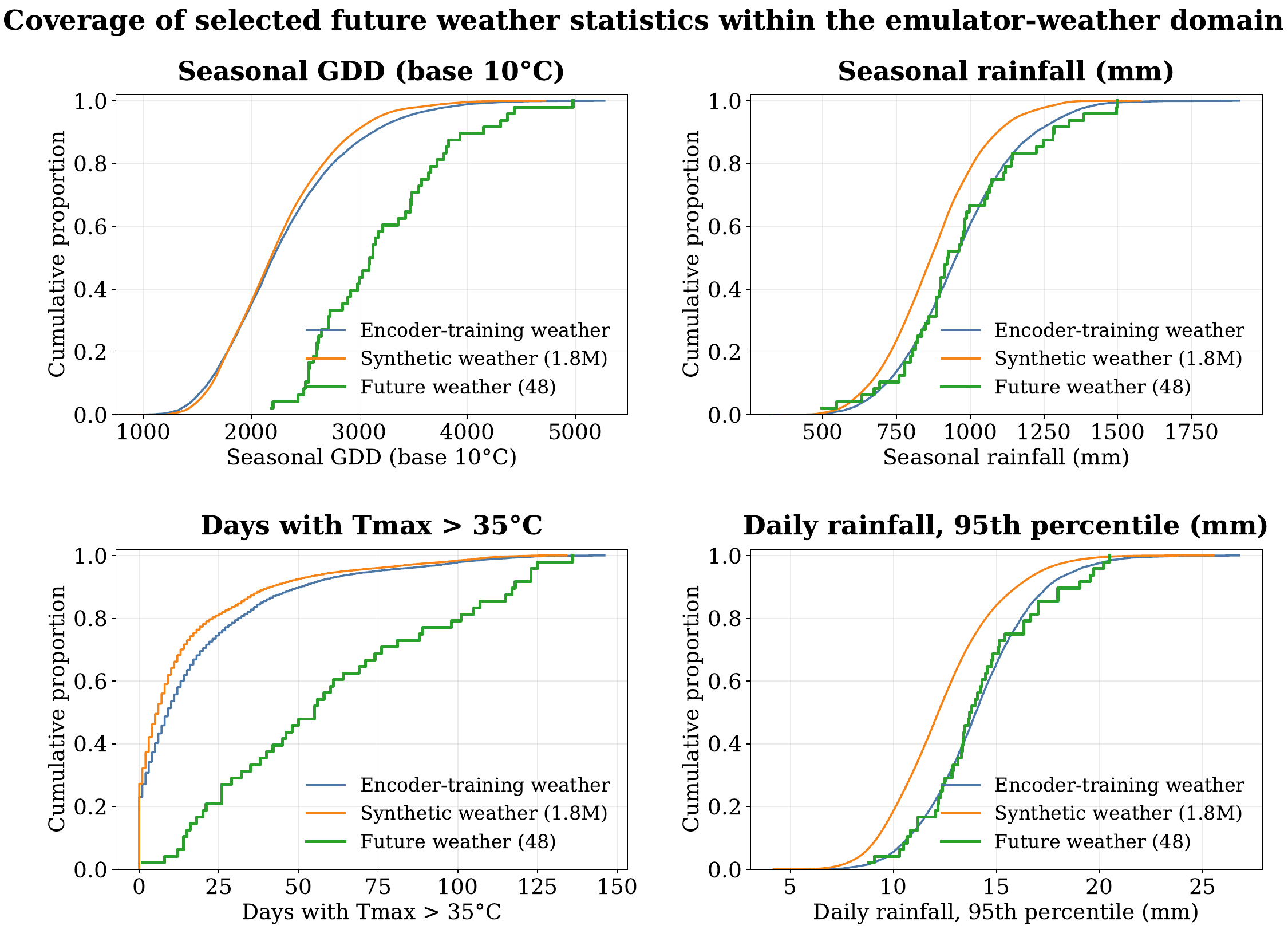}
    \captionsetup{width=1\textwidth}
    \caption{Coverage of selected future-weather statistics within the emulator-weather domain. Empirical cumulative distribution functions compare encoder-training weather (blue), 1.8 million convexly sampled synthetic weather sequences (orange), and the 48 future location--climate-model--scenario weather files used as year-2100 cases (green). The gray background used in an earlier layout represented the 1st--99th percentile range of the encoder-training weather; it is omitted here to keep the comparison focused on the three distributions.}
    \label{fig:selected_weather_ecdf_coverage}
\end{figure}

\bibliography{sn-bibliography}